# Wavelength and Polarization Dependence of Second Harmonic Responses from Gold Nanocrescent Arrays

Hiroaki Maekawa,[1] Elena Drobnyh,[2] Cady A. Lancaster,[3] Nicolas Large,[4] George C. Schatz,[5] Jennifer S. Shumaker-Parry,[3] Maxim Sukharev,[2,6,*] and Nien-Hui Ge[1,*]

[1] Department of Chemistry, University of California at Irvine, Irvine, California 92697-2015, USA

[2] Department of Physics, Arizona State University, Tempe, Arizona 85287, USA

[3] Department of Chemistry, University of Utah, Salt Lake City, Utah 84112, USA

[4] Department of Physics and Astronomy, University of Texas at San Antonio, San Antonio, Texas 78249, USA

[5] Department of Chemistry, Northwestern University, Evanston, Illinois 60208-3113, USA

[6] College of Integrative Sciences and Arts, Arizona State University, Mesa, Arizona 85212, USA

\* To whom correspondence should be addressed.

Maxim Sukharev: Maxim.Sukharev@asu.edu

Nien-Hui Ge: nhge@uci.edu

**Abstract**

In the developing field of nonlinear plasmonics, it is important to understand the fundamental relationship between properties of the localized surface plasmon resonance (LSPR) of metallic nanostructures and their nonlinear optical responses. A detailed understanding of nonlinear responses from nanostructures with well characterized LSPRs is an essential prerequisite for the future design of sophisticated plasmonic systems with advanced functions to control light. In this article, we investigate the second-order harmonic (SH) responses from gold nanocrescent (Au NC) antennas which have wavelength and polarization sensitive LSPRs in the visible and near-infrared wavelength range. The wavelength dependence of the SH intensity exhibits spectral profiles different from dipole LSPR bands in absorbance spectra. The incident polarization angle dependence was found to vary significantly when the excitation wavelength was tuned over the dipole band. Finite-difference time-domain calculations coupled with a nonlinear hydrodynamic model were carried out for Au NC arrays to investigate the local field enhancement of the incoming fundamental and emitting SH light. The experimental and theoretical results indicate that the effects of higher-order LSPRs, such as quadrupole and multipole resonances, occurring at SH wavelengths are important in governing the SH generation process. Also, it is shown that the incident polarization angle dependence of SH signals is very strongly sensitive to nanoscale variations in the NC's shape.

**I. Introduction**

The local field enhancement caused by a localized surface plasmon resonance (LSPR), that is, a coherent oscillation of electrons in the conduction band of metal nanostructures, has opened up several new research realms. Among them, nonlinear plasmonics has attracted much attention because of its broad applications, ranging from the development of new spectroscopic methods, to fabrication of metamaterials with unique functions.[1-4] An underlying and common concept is to utilize LSPR to induce large nonlinear polarization and to harness the resulting nonlinear optical phenomena. Beyond roughened metal surfaces with nanoscale islands, much effort has been made to fabricate plasmonic nanostructures that can control and manipulate light in a well-controlled manner. At the same time, it is essential to characterize the LSPR and its effects on nonlinear optical properties for better understanding and further improvement of nonlinear plasmonic devices. To this end, nonlinear optical spectroscopic techniques, especially second-harmonic (SH) generation,[5-33] third-harmonic generation,[34-36] optical Kerr effect,[37,38] and four-wave mixing,[39,40] have been applied to many kinds of plasmonic structures.

SH responses from a variety of metal nanoantennas have been vigorously investigated, including nanoparticles,[8-12] nanocylinders,[13] split-ring resonators,[14-16] nanorods,[17] G-shaped,[18] L-shaped,[19-22], T-shaped,[23] and V-shaped antennas,[24] as well as more complicated nanostructures.[7,25-27] Theoretical approaches also have been developed to explain the mechanism of SH generation and to predict the nonlinear optical responses by numerical calculations.[28-33] The key point in these experimental and theoretical studies was to reveal how strongly the local fields are enhanced by LSPR excitation. This information is important because the concept of phase matching used in nonlinear optics for bulk materials is invalid for nanostructures with dimensions much smaller than the wavelength of light. Also, there is significant interest in detecting nonlinear signals from single nanostructures, which would only be possible if very large nonlinear polarizations can be induced via field enhancement. After the initial studies that examined the effects coming from enhancing local field excitation, further studies have been carried out on double-resonance plasmonic nanoantennas to increase the efficiency of SH frequency conversion.[41-43] Field enhancement by LSPR excitation occurring at the SH wavelength was indeed experimentally confirmed by some groups.[17,24] However, it also was reported that LSPR excitation at the SH wavelength just reabsorbs the generated SH field and does not contribute to further enhancement of the emitted field beyond the excitation field enhancement.[14] This contradiction of prior findings suggests that further investigations are needed. Tuning of LSPRs in a single type of structure could provide more understanding of the fundamental processes involved in SH responses from plasmonic nanostructures. In complex nanoantennas, double resonance may be achieved by involving not only dipole LSPR, but also higher-order LSPRs, such as quadrupole and octupole resonances.

In this work, we performed experimental and theoretical studies of SH response from two gold nanocrescent (Au NC) arrays with distinct LSPRs. Previously, it has been shown that metallic NC structures exhibit LSPRs that can be selectively excited by varying the wavelength and polarization of incident light.[44-47] The peak wavelength of resonance bands can be tuned by changing the dimensions of the NC. This feature makes it feasible to have, for example, a dipole LSPR band in the near-infrared (IR) range for enhancement of the excitation field and a higher-order LSPR band in the visible range for enhancement of the SH field. In SH spectroscopy experiments, we measured the wavelength as well as polarization dependence of signals from the NC arrays. The latter is especially sensitive to the source of SH polarization.[9] Comparisons of the polarization dependence at different excitation wavelengths over the dipole band will provide information on the roles of higher-order LSPRs in the SH generation of Au NC antennas. If only the dipole LSPRs at the fundamental wavelength are important, then we expect to observe the same polarization dependence regardless of the excitation wavelength. On the contrary, if higher-order LSPRs are involved in the SH generation process, the polarization dependence of SH intensity will exhibit dependence on the excitation wavelength. We applied a theoretical model to explain the experimental findings by using a macroscopic framework based on NC symmetry. We also performed finite-difference time-domain (FDTD) calculations coupled with a nonlinear hydrodynamic model to directly evaluate SH fields generated from NC arrays. This includes studies of the wavelength and polarization dependence of the SH response, along with the effects of higher-order LSPRs.

## II. Experimental Section

**A. Fabrication and characterization of nanocrescent arrays.** Nanosphere template lithography was used to fabricate Au NCs and has been described previously.[44,46,48] The templates used were 220 nm or 505 nm sulfate/aldehyde functionalized polystyrene nanospheres (Life Sciences Solutions, Inc., Grand Island, NY) that were closed-packed onto cleaned (see references) 1″ x 1″ glass slides (BK7 glass, Ted Pella Inc. Redding, CA) using a method adapted from that used by Weiss and coworkers.[49,50] The polystyrene template arrays were etched by exposure to oxygen plasma. Substrates were placed at an angle of 40° with respect to the metal source in an electron beam evaporator chamber (Denton SJ20C Vacuum USA, Moorestown, NJ). Gold (25 nm film thickness) was deposited with a rate of 1 Å/s as measured by a quartz crystal microbalance (XPC2 Inficon, East Syracuse, NY) under high vacuum ($2.0 \times 10^{-6}$ mTorr). Argon ion milling (PlasmaLab 80 Plus, Oxford Instruments) was used to remove the Au film at a power of 100 W and flow rate of 10 sccm. After milling, the polystyrene beads were removed by lift-off using transparent tape and samples were stored under nitrogen.

Atomic layer deposition (Cambridge NanoTech Fiji F200 Plasma, Cambridge, MA) was used to deposit alumina films on the Au NC substrates to form alumina-coated Au NCs ($Al_2O_3$-Au NC).

Trimethylaluminum (TMA) and water pulsed alternately in a nitrogen carrier stream using a growth temperature of 33 °C and base pressure of $1.6 \times 10^{-6}$ Torr. The process occurred in four steps: (1) 0.06 s pulse of water, (2) 60 s purge with nitrogen, (3) 0.06 s pulse of TMA, and (4) 10 s purge with nitrogen. Each cycle deposited ~0.1 nm of material.[51]

Scanning electron microscope (SEM) images of the fabricated samples are in Figure 1a,b. On the array sample prepared with the template bead size of 220 nm, NCs have an average tip-to-tip length $L$ = 220 nm and average width $w$ = 85 nm (hereafter this sample is called NC1). Another array sample (NC2) prepared with the 505 nm-template beads has NCs with the average sizes of $L$ = 430 nm and $w$ = 90 nm.

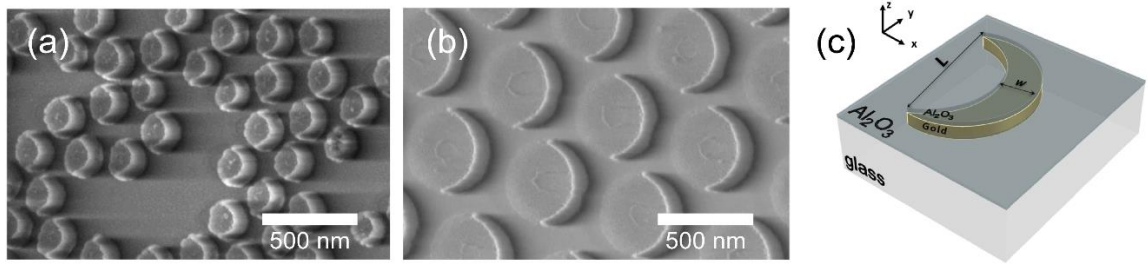

**Figure 1.** SEM images of Au nanocrescent (NC) array samples: (a) NC1; (b) NC2 (see Section II.A. for fabrication details and dimensions). Schematic diagram in (c) shows the unit cell used for numerical simulations. The tip-to-tip length and width of NC is represented by $L$ and $w$, respectively. Two sets of $L$ and $w$ were used to simulate two NC arrays (NC1′: $L$ = 220 nm, $w$ = 85 nm; NC2′: $L$ = 430 nm, $w$ = 90 nm).

**B. Visible / near-IR spectrum measurement.** Visible/near-IR absorbance spectra of NC1 and NC2 were recorded by a spectrometer equipped with a control module for incident light polarization (PerkinElmer, Lambda 750). We measured the polarization dependent absorbance at a fixed wavelength at 1300 nm for NC1 and 1277 nm for NC2, and fitted the data with the cosine squared of the incident polarization angle to optically determine the orientation of NCs for each sample.

**C. SH spectroscopy measurement.** SH responses from the NC samples were measured using a home-built SH spectrometer (Figure S1). A home-built optical parametric amplifier (OPA) pumped by the output 1-kHz, 70-fs Ti:Sapphire regenerative amplifier (Spectra-Physics, Spitfire ACE) generates near-IR pulses tunable from 1200 to 1600 nm. The near-IR spectrum has a Gaussian shape with a spectral width of about 15–20 nm over the tunable wavelength range. The polarization of the incident near-IR pulse was controlled by an achromatic half-wave plate. A long-pass filter removed SH light generated by any optics in the OPA and the half-wave plate so that only SH responses from the NC samples were measured. The incident light was loosely focused onto the sample surface with a gold-coated spherical concave mirror with a focal length of 100 mm. The incident power was decreased to about 0.5 mW to avoid any damage to the Au NC during measurements. SH light generated from the NC array passed through the glass substrate and was collimated with another concave mirror. After removing the residual near-IR excitation pulses using a short-pass filter, an achromatic linear polarizer was used as an analyzer to select the polarization component of the SH signal, which is either along the short axis (SA) or long axis (LA) of the NC. An achromatic half-wave plate placed after the analyzer rotates the polarization to the same direction in front of the monochromator to eliminate the polarization dependence of grating diffraction efficiency. The SH signal was dispersed in the monochromator, and its spectrum was acquired with an electron multiplying charge coupled device camera cooled to −70 °C. Wavelength calibration of the spectrograph was carried out using mercury/argon and neon lamps. To investigate the incident wavelength dependence of SH intensity from the NC samples, the center wavelength of the near-IR pulses was tuned from 1200 to 1600 nm with a step of 25 nm. For such a wide tuning range of wavelength, it is indispensable to carefully calibrate or eliminate the wavelength dependence of the light source, such as intensity and pulse duration, and optics in the spectrometer setup. To this end, right after measuring the SH signal from the sample, we used two flipper mirrors to steer the near-IR light to another path and measured the SH response of a 5-mm thick z-cut quartz plate under the same experimental condition at each wavelength. The second-order nonlinear susceptibilities of quartz, as estimated from its dielectric functions, are almost invariant over the tuned wavelength, which is consistent with some reported values in the literature.[52] The Fresnel factors for light impinging on an air/quartz interface with an incident angle of 45° are very insensitive to wavelength. The wavelength dependence of SH response of the NC samples was obtained by dividing the integrated SH spectral intensity from the NC by that from z-cut quartz.

**III. Theory**

**A. Calculation under macroscopic framework.** For an accurate calculation of SH response from metal plasmonic nanostructures, it is necessary to take into account the local enhancement factors for the incident fundamental and generated SH field, and their variations over the nanometer scale structures. Therefore, the conventional method for calculating the nonlinear polarization of macroscopic systems is not quite applicable to Au NC arrays. However, it is still useful to predict the SH response and polarization dependence based on the object symmetry. This macroscopic prediction will allow us to qualitatively discuss how microscopic effects alter the observed signals.

The surface symmetry of the NC is $C_{1h}$, and it is placed on the *x-y* plane of the Cartesian coordinate system with the LA along the *y*-axis, as shown in Figure 2. Under the electric dipole approximation, and when the incident light is propagating normal to the *x-y* plane, the nonvanishing second-order nonlinear susceptibility elements that need to be included are $\chi^{(2)}_{xxx}$, $\chi^{(2)}_{xyy}$, and $\chi^{(2)}_{yxy} = \chi^{(2)}_{yyx}$.[53] The induced second-order nonlinear polarization $P^{(2)}_i(2\omega)$ along the $i$-axis ($i = x$ or $y$) is described as follows:

$$P^{(2)}_x(2\omega) \propto \chi^{(2)}_{xxx} \cos^2\theta + \chi^{(2)}_{xyy} \sin^2\theta \qquad (1a)$$

$$P^{(2)}_y(2\omega) \propto \chi^{(2)}_{yxy} \sin\theta \cos\theta \qquad (1b)$$

where $\theta$ is the angle between the $x$-axis and the polarization direction of the incident light. Figure 2 shows polar plots of the induced SH intensity as a function of the polarization angle of the incident beam. When the polarization of the induced SH beam is along the $x$-axis (SA of NC), the polar plot pattern (Figure 2a–c) significantly depends on the ratio between the two susceptibility tensors, $\chi^{(2)}_{xxx}$ and $\chi^{(2)}_{xyy}$. In the limit of $\left|\chi^{(2)}_{xxx}\right| \gg \left|\chi^{(2)}_{xyy}\right|$, a two-lobe pattern oriented along the $x$-axis results (Figure 2a). The shape of this polar pattern changes from the two-lobe to a four-lobe along the x- and y-axes for $\left|\chi^{(2)}_{xxx}\right| = \left|\chi^{(2)}_{xyy}\right|$ (Figure 2b), and shows a two-lobe pattern oriented along the $y$-axis in the limit of $\left|\chi^{(2)}_{xxx}\right| \ll \left|\chi^{(2)}_{xyy}\right|$ (Figure 2c). In contrast, when the polarization of the induced SH beam is along the $y$-axis (LA of NC), eq 1b indicates that $\left|P^{(2)}_y\right|^2$ exhibits the same $\theta$ dependence regardless of the value of $\chi^{(2)}_{yxy}$, which is a four-lobe pattern oriented along the diagonal and antidiagonal on the $x$-$y$ plane (Figure 2d).

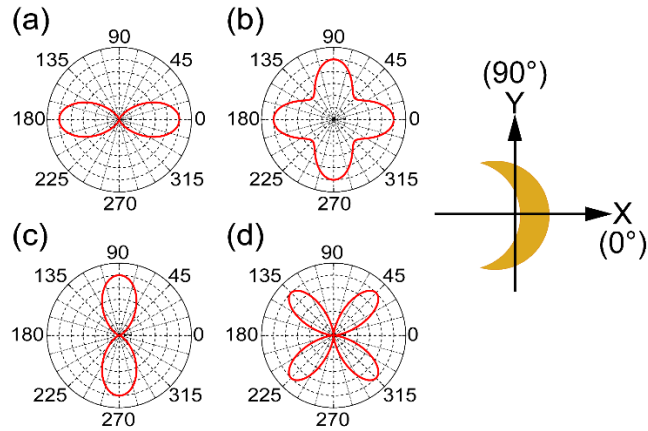

**Figure 2.** Polar plots of SH intensity as a function of the polarization angle of the incident beam, calculated under the electric dipole approximation. The polarization of SH is along the short axis (SA) of the NC for parts (a–c) and along the long axis (LA) for part (d). The ratio of second order susceptibility elements $\left|\chi^{(2)}_{xxx}\right|/\left|\chi^{(2)}_{xyy}\right| = 10$ (a); 1 (b); 0.1 (c), and the phase difference between the two elements was set to $\pi/2$.

Previous studies analyzed the incident polarization angle dependence of SH responses[19,20] from metallic nanostructures by the equation below

$$P_i^{(2)}(2\omega) \propto \xi_{ixx}^{(2)} \cos^2\theta + \xi_{ixy}^{(2)} \sin\theta \cos\theta + \xi_{iyy}^{(2)} \sin^2\theta \quad (i = x, y) \quad (2)$$

which includes forbidden terms under the electric dipole approximation. The same functional form has also been used to fit the polarization dependent hyper Rayleigh scattering (HRS) intensities from nanoparticles.[8-10,12] The additional terms are required to include nonlinear polarization effects induced by quadrupole and magnetic dipole interactions. We used eq 2 to fit our experimental data because some of the measured polarization dependence could not be reproduced by the symmetric two- or four-lobe patterns from eqs 1a and 1b.

**B. Calculation based on the hydrodynamic model.** The dynamics of electromagnetic radiation at dielectric-plasmonic interfaces follows Maxwell's equations

$$\varepsilon_0 \frac{\partial \vec{E}}{\partial t} = \frac{1}{\mu_0}\nabla \times \vec{B} - \frac{\partial \vec{P}}{\partial t},$$

$$\frac{\partial \vec{B}}{\partial t} = -\nabla \times \vec{E}, \quad (3)$$

where the macroscopic polarization, $\vec{P}$, is described by the hydrodynamic model for the conduction electrons.[54,55] The equations of motion governing the dynamics of conduction electrons are defined by spatio-temporal variations of the electron velocity field and the electron number density resulting in the following partial differential equation[56]

$$\frac{\partial^2 \vec{P}}{\partial t^2} + \gamma \frac{\partial \vec{P}}{\partial t} = \frac{n_0 e^2}{m^*}\vec{E} + \frac{e}{m^*}\frac{\partial \vec{P}}{\partial t} \times \vec{B} - \frac{e}{m^*}\vec{E}(\nabla \cdot \frac{\partial \vec{P}}{\partial t}) - \frac{1}{n_0 e}[(\nabla \cdot \frac{\partial \vec{P}}{\partial t})\frac{\partial \vec{P}}{\partial t} + (\frac{\partial \vec{P}}{\partial t} \cdot \nabla)\frac{\partial \vec{P}}{\partial t}]. \quad (4)$$

Here $n_0$ is the equilibrium number density of conduction electrons, $m^*$ is the effective electron mass, and $\gamma$ is the phenomenological decay constant.

Equations 3 and 4 are coupled and numerically propagated in space and time using home-built finite-difference time-domain codes. The numerical details of the propagation scheme and parallel spatial decomposition can be found elsewhere.[57] In our calculations the numerical convergence for both linear and nonlinear responses is achieved for the spatial resolution of 1.5 nm with the time step of 2.5 as. When simulating second harmonic response the system is excited by a 200-fs laser pulse with a peak amplitude of $2 \times 10^{-2}$ V/nm. We found that the angular properties of the second harmonic signal are highly dependent on the duration of the pump pulse converging for pulses longer than 180 fs. To avoid mixing different nonlinear processes the pump peak amplitude is chosen to ensure simulations are in the perturbative regime, i.e., the second harmonic signal scales quadratically with the pump intensity. The total propagation time of the coupled hydrodynamic-Maxwell's equations is 500 fs. The following numerical parameters in eq 4 are used to describe gold: $n_0 = 5.9 \times 10^{28}$ m$^{-3}$, plasma frequency $\Omega_p = \sqrt{\frac{n_0 e^2}{\varepsilon_0 m^*}} = 1.17 \times 10^{16}$ rad/sec (7.71 eV), and $\gamma = 8.23 \times 10^{-2}$ eV. These parameters are tuned such that the corresponding dielectric constant obtained from the Drude model (the linear limit of eq 4) match the experimental data at 1200 nm.[58] We consider a periodic array of Au NCs schematically depicted in Figure 1c, with two different sets of $L$ and $w$ (NC1′: $L$ = 220 nm, $w$ = 85 nm; NC2′: $L$ = 430 nm, $w$ = 90 nm). The nanoparticle is placed on top of a semi-infinite nondispersive dielectric slab with refractive index $n$ = 1.456. The input side of the array is covered by a thin layer of Al$_2$O$_3$ (modeled as a dielectric with refractive index of 1.765). All simulations are performed at AFRL/DSRC HPC clusters Mustang and Onyx using 1584 processors. Typical execution times of our codes vary between 10 (NC1′) – 20 (NC2′) minutes for linear simulations and 80 (NC1′) – 90 (NC2′) minutes to obtain SHG results.

## IV. Results and Discussion

**A. Polarization dependent LSPRs of NC array.** Parts a and b of Figure 3 show the visible/near-IR absorbance (the decadic logarithm of the inverse transmittance) spectra of NC1 and NC2, respectively, recorded with the incident light polarized along the LA (red) and SA (blue) of the NCs. The width of LSPR bands is much larger than that of the excitation pulse spectrum (shown as the dotted line in Figure 3a). The obvious polarization dependence of absorbance spectra indicates that the wavelength and polarization of the incident light both play an important role to selectively excite a specific NC plasmon resonance band. Previously, Shumaker-Parry and coworkers studied the optical properties of NC arrays with the diameter ranging from 100 to 500 nm, and assigned the characteristic resonance modes with the aid of FDTD calculations.[46] Here we assign the LSPR modes based on this assignment and the electric field distribution calculated in our study (shown in Figures S4–S7, S12, S14–18). NC1 exhibits peaks at 1300 and 684 nm for the light polarized along the LA, originating from the dipole and quadrupole resonance, respectively. When the light is polarized along the SA, the dipole band is located at 840 nm. The peaks at around 550–560 nm are insensitive to polarization and were assigned to out-of-plane dipole modes in the previous study.[46] However, our FDTD calculation for NC1′ showed a multipole pattern (Figure S14) for this band. The identification of multipole plasmon modes in this wavelength region is in agreement with recent experimental and theoretical orientation-dependent analysis of plasmonic and chiral plasmonic responses of gold NCs.[59] For NC2, which is an array of NC larger than that in NC1, the dipole and quadrupole bands are redshifted. The LA dipole and quadrupole resonance bands have the peak wavelengths at 2190 and 1001 nm, respectively. The SA dipole band is redshifted to 1277 nm. The resonance band centered at 560–580 nm, exhibits very marginal wavelength shift between NC1 and NC2 compared to the dipole and quadrupole bands. The peak wavelengths of the LSPR modes are summarized in Table 1.

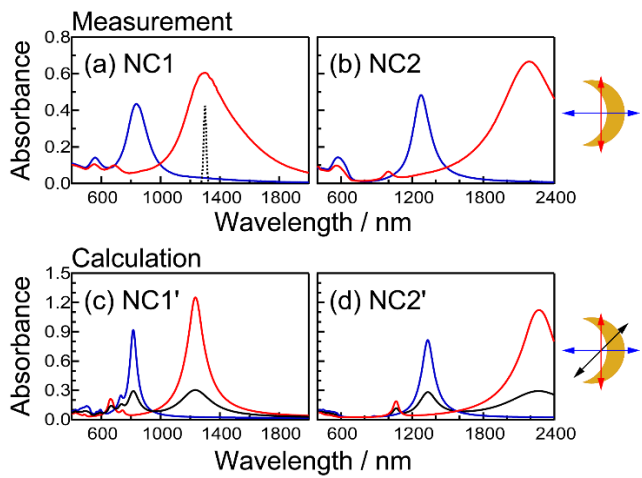

**Figure 3.** Polarization dependent visible/near-infrared absorbance spectra measured for NC1 (a) and NC2 (b), and calculated for NC1′ (c) and NC2′ (d). Red, blue and black curves represent spectra acquired with the incident light polarized along the long axis (LA), short axis (SA), and at 45° from the two axes of nanocrescents, respectively. The black dashed line in (a) shows the excitation pulse spectrum used in the SH spectroscopy measurements when the near-IR output is tuned to 1300 nm.

**Table 1.** The peak wavelengths (in nm) of localized surface plasmon resonance modes measured (NC1, NC2) and calculated (NC1′, NC2′) for gold nanocrescent arrays.

|  | Dipole | | Quadrupole | Multipole[a] | |
|---|---|---|---|---|---|
|  | LA | SA | LA | LA | SA |
| NC1 | 1300 | 840 | 684 | 552 | 558 |
| NC1′ | 1237 | 815 | 661 | – | – |
| NC2 | 2190 | 1277 | 1001 | 563 | 577 |
| NC2′ | 2270 | 1333 | 1065 | – | – |

[a]The peak wavelengths of NC1′ and NC2′ were not determined due to the calculated spectral shape. See Figure 3c,d.

**B. Calculated absorbance spectra of Au NC.** Linear absorbance calculated at three incident polarizations is shown in Figure 3c,d as a function of the incident wavelength. The peak wavelengths of simulated LA and SA dipole LSPR are listed in Table 1. Given the fact that linear simulations rely on an ideal square lattice for both NC1′ and NC2′ and are based on a simple Drude model, the agreement with experimental results for the dipole plasmon modes is remarkable. We note that if the period used in calculations is set too short, SA and LA modes are significantly altered. We thus performed a set of simulations with different periods in order to find the shortest period for each geometry, at which plasmon modes are nearly independent on small period variations. The value of the "sweet" period is found to be at 360 nm and 576 nm for NC1′ and NC2′, respectively.

**C. Polarization dependence of SH response from NC arrays.** Figure S2a shows the SH spectra measured with the peak wavelength of incident light tuned to the peak of LA (SA) dipole mode for NC1 (NC2). The polarization direction of the excitation beam is along the LA (SA) for NC1 (NC2). The polarization direction of SH light is along the SA for both samples. The measured SH spectra have a FWHM of ~10 nm and exhibit a Gaussian shape, similar to the shape of the incident near-IR pulses.

To verify the order of the nonlinear process, we measured SH intensities from the two NC samples and the z-cut quartz as a function of the near-IR excitation power and plotted them in a double logarithmic scale in Figure S2b. The experimental data and linear least square fitting clearly indicate that the observed SH intensities are proportional to the square of incident light intensity.

Parts a and b of Figure 4 show the polarization dependence of SH response from NC1 and NC2, respectively. The angle of polar plots indicates the polarization direction of the incident near-IR pulse, which is defined as 0° (180°) when the light is polarized along the SA of NC and 90° (270°) when polarized along the LA. The polarization component of the detected SH signal is either along the SA (top row) or the LA (bottom row). The incident near-IR wavelength was centered at 1250, 1300, 1380, and 1550 nm, which covered the LA dipole resonance band for NC1; and 1280, 1400 and 1540 nm, covering the SA dipole resonance band for NC2. The observed SH intensities significantly depended on the samples, wavelength, and incident and detected polarization direction.

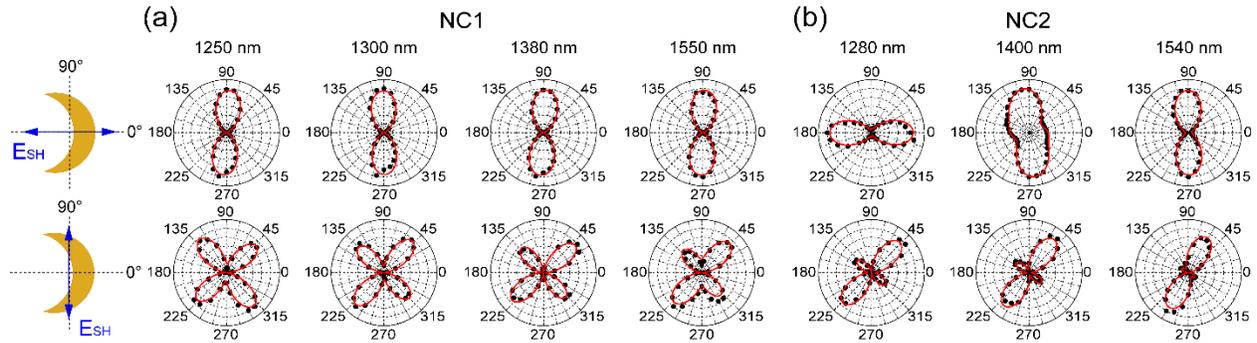

**Figure 4.** Incident polarization angle dependence of second harmonic (SH) signals measured for NC1 (a) and NC2 (b) at different excitation wavelengths. The plots have been normalized to the maximal intensity. The angle is defined as 0° (90°) when the excitation pulse is polarized along the SA (LA) axis of the NC. Blue arrows indicate the polarization direction of detected SH signals. Black dots represent experimental data acquired with the polarization angle varying from 0° to 360° with a step of 10°. Fitting results using eq 2 are overlaid in the polar plots as red solid lines.

The polar plots of SH intensities along the SA of NC1 (Figure 4a, top) show a two-lobe pattern, regardless of the incident wavelength. The intensity is strongest when the incident light is polarized along the LA of the NC, and it is decreased by two orders of magnitude when incident polarization is along the SA. When the analyzer angle is set to 90° (Figure 4a, bottom), the polar plots exhibit a four-lobe pattern. The amplitudes of the lobes at the antidiagonal (135° and 315°) slightly depend on the incident wavelength. Equation 1a based on the macroscopic formalism can be used to reproduce the two-lobe patterns observed at the different wavelengths. However, the asymmetric four-lobe pattern, especially pronounced at 1550 nm, suggests that this equation is not adequate to describe the SH responses of plasmonic NC antenna.

More drastic wavelength dependence of the SH intensity was observed for NC2, especially when the SH component along the NC SA was recorded (Figure 4b, top). A two-lobe pattern is oriented along the SA at the incident wavelength of 1280 nm, indicating that the SH intensity became strongest (weakest) for the horizontally (vertically) polarized excitation light. In contrast, the orientation of the two-lobe pattern is along the LA at 1540 nm. When the polarization dependence was measured at 1400 nm, almost the middle of the two wavelengths, we observed a distorted polar plot pattern, unlike the clean two lobes as seen at 1280 and 1540 nm. The SH polarized along the LA exhibits a four-lobe pattern, but the amplitudes of two lobes along the diagonal are larger than those along antidiagonal (Figure 4b, bottom). The ratio of antidiagonal to diagonal lobe size decreases with increasing the excitation wavelength. Also, the orientation of the larger two-lobe changes from 45° at 1280 nm to 60° at 1540 nm.

It is worth examining whether the polarization dependence of SH intensity is affected by the NC orientation. We rotated the samples by 180° on the sample plane so that the incident light remained focused on the air/NC surface first, and the generated SH signal emitted to the far-field through the glass substrate of the sample. Figure S3 shows the experimental results observed for NC2 at the excitation wavelength of 1280 nm. The symmetric two-lobe pattern and asymmetric four-lobe pattern are the same for the two NC orientations.

As the absorbance spectra in Figure 3a shows, we can excite the dipole LSPR band of NC1 selectively by setting the polarization of incident light to the NC LA in the wavelength range from 1200 to 1600 nm. In the simplest model of this result, the field enhancement by LSPR can be taken into account in calculating nonlinear polarizations by incorporating the enhancement factor $L_i(\omega)$ ($i = x$ or $y$).[17,26,60] The effective radiating second-order nonlinear polarization along the NC SA ($x$-axis, eq 1a) can be written as

$$P_{x,\text{eff}}^{(2)}(2\omega) = L_x(2\omega)P_x^{(2)}(2\omega) \propto L_x(2\omega)\left[\chi_{xxx}^{(2)}L_x(\omega)L_x(\omega)\cos^2\theta + \chi_{xyy}^{(2)}L_y(\omega)L_y(\omega)\sin^2\theta\right]$$
$$\equiv \tilde{\chi}_{xxx}^{(2)}\cos^2\theta + \tilde{\chi}_{xyy}^{(2)}\sin^2\theta \qquad (5)$$

For NC1, on the basis of the absorbance spectrum, it is inferred that $L_y(\omega) \gg L_x(\omega)$ in the excitation wavelength range, and hence $\tilde{\chi}_{xyy}^{(2)} \gg \tilde{\chi}_{xxx}^{(2)}$ as long as the value of $\chi_{xyy}^{(2)}$ is not much smaller than that of $\chi_{xxx}^{(2)}$. Interestingly, as shown in a previous study of aluminum rod-type nanoantennas, it is a good approximation to assume equal absolute values of $\chi_{ppp}^{(2)}$ and $\chi_{spp}^{(2)}$, where s and p represent the polarization along the short and long side of nanorods.[17] Even if we assume the values of $\chi_{xxx}^{(2)}$ and $\chi_{xyy}^{(2)}$ to be similar, the condition that $\tilde{\chi}_{xyy}^{(2)} \gg \tilde{\chi}_{xxx}^{(2)}$ still holds. This condition corresponds to the limit where the SH intensity polar plot exhibits a two-lobe pattern along the y-axis (Figure 2c). We indeed observed very similar two-lobe patterns at four different excitation wavelengths (Figure 4a), and hence the effective susceptibility $\tilde{\chi}_{xyy}^{(2)}$ dominates the SH response for this specific polarization.

For NC2, the polarization dependence of the SH intensity along the SA can be similarly explained by eq 5. At 1280 nm, which is very close to the peak of the SA dipole resonance, it is expected that the incident field along the x-axis is enhanced, and hence $L_x(\omega)$ is much larger than $L_y(\omega)$. This gives rise to the limit of a two-lobe pattern along the x-axis in the polar plot of the SH polarization dependence (Figure 2a), which is consistent with the experimental result (Figure 4b). It is possible that $L_y(\omega)$ becomes much larger than $L_x(\omega)$ at 1540 nm because NC2 exhibits a higher absorbance when the polarization is along the NC LA than along the SA at this wavelength. In this case, we can follow the same logic as mentioned for NC1, and thereby explain the two-lobes along the y-axis (LA) as observed. However, the polarization dependence at 1400 nm deviates from eq 5, and hence including $L(\omega)$ is not sufficient to explain the underlying plasmon effects on the SH generation.

Similarly, eq 1b can be rewritten to include the enhancement factors to become:

$$P_{y,\text{eff}}^{(2)}(2\omega) = L_y(2\omega)P_y^{(2)}(2\omega) \propto L_y(2\omega)\chi_{yxy}^{(2)}L_x(\omega)L_y(\omega)\sin\theta\cos\theta$$
$$\equiv \tilde{\chi}_{yxy}^{(2)}\sin\theta\cos\theta \quad (6)$$

Theoretically, as indicated by eq 6, the enhancement factor does not affect the polarization dependence of the SH intensity detected along the LA. For NC1, although the polarization dependence measured at 1250 nm was reasonably reproduced by this functional form, the amplitudes of the four-lobe pattern become unequal with increasing the excitation wavelength. This distortion may indicate effects of LSPR excitation other than enhancement of the fundamental field. In previous studies, contributions of electric dipole and quadrupole plasmon resonances to HRS signals from silver nanoparticles were distinguished by tuning the SH wavelength to specific plasmon bands.[9,61] The polarization dependence of HRS intensity also showed different polar patterns depending on wavelength.[9] Therefore, the increased asymmetry in the measured SH polar plots with wavelength suggest that dipole-forbidden higher order LSPRs may play a role here. For NC2, the polar plots of SH intensity along the LA are even more asymmetric than those of NC1. The unequal amplitudes of the diagonal and antidiagonal lobes, and the rotating patterns with excitation wavelength cannot be explained under the electric dipole approximation.

The observed polarization dependence was fit better by eq 2 than eqs 1a and 1b, and the resulting polar patterns are overlaid on the same graphs (Figure 4a,b, red lines). Resultant fitting coefficients $\xi^{(2)}_{ijk}$ for the different polarization components are plotted in Figure 5. The two-lobe patterns for the SH intensity polarized along the SA of NC1 are mainly dictated by the *xyy* term with tiny contributions from *xxx* and *xxy*. The situation is the same for the SH intensity along the SA of NC2 only at 1540 nm. The contribution of *xxx* increases with decreasing the wavelength, and at 1280 nm, it almost exclusively dominates the observed SH response. The four-lobe patterns, acquired for the SH intensity along the NC LA, are mainly caused by the dipole-allowed *yxy* component. The other two dipole forbidden terms, *yyy* and *yxx*, distort the symmetry of the four-lobe pattern, and their contributions depend on the incident light wavelength. The fitting results suggest that dipole allowed polarization tensors are the main contributors to the SH response for both samples. However, dipole forbidden components for $C_{1h}$ symmetry are also important, especially for the SH intensity along the LA. Take NC2 as an example, if $\xi^{(2)}_{yyy}$ makes a contribution to the observed SH signal, it would be increased by $L_y(\omega)^2$ as the incident wavelength is tuned toward the peak of LA dipole resonance, and hence the contribution of a two-lobe pattern at 90° to the polar plot would increase. The observed pattern for NC2 at 1540 nm (Figure 4b) may be a manifestation of this transition between a four-lobe and two-lobe pattern, and the SH intensity is strongest at 60°.

Another possible reason for the asymmetric four-lobe patterns may be related to a marginally distorted NC shape as discussed below with numerical calculation results.

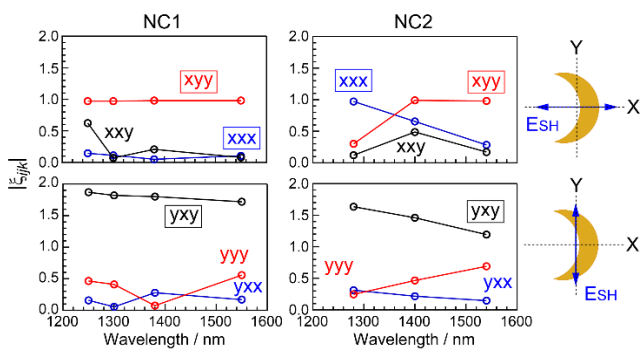

**Figure 5.** Fitting coefficients $\xi^{(2)}_{ijk}$ in eq 2 to reproduce the incident polarization angle dependence of SH signals in Figure 4 for NC1 (left) and NC2 (right) at different excitation wavelengths. The polarization direction of the detected SH signals is along the SA (top) or LA (bottom) of the NC. Polarization components, *ijk*, are indicated by the labels. Polarization components that are allowed under the dipole approximation for $C_{1h}$ symmetry are marked with rectangular boxes.

**D. Calculated polarization dependence of SH signal.** To put the nonlinear model to the test, we performed simulations of SHG for both NC1′ and NC2′. The SH electric field distributions with the excitation at peak wavelengths of dipole and quadrupole LSPRs are shown in Figures S8–S11 and S13, respectively. Moreover, we investigated the polarization dependence of the SH signal by calculating power spectra at a given pump wavelength while varying the in-plane incident polarization angle. When performing nonlinear simulations, we separately evaluated SH polarized along SA and LA. The results are shown in Figure 6a,b, where we examined characteristic wavelengths: the peak wavelengths of SA/LA mode and the wavelength at which linear absorbances calculated for the incident polarization along SA and LA intersect (Figure 3c,d). Firstly, we note that the SH signal polarized along LA has four lobes as in the experiments. Secondly, as expected the SA polarized signal is oriented either along SA or LA depending on which plasmonic mode is dominant at a given pump wavelength. These results reproduce expected angular properties of SH as obtained from the second-order nonlinear polarization as in eqs 1a and 1b. Thirdly, the four-lobe pattern oriented along SA and LA is seen for both NC1′ and NC2′ at wavelengths for which both plasmonic modes are being pumped (943 nm for NC1′ and 1560 nm for NC2′). This clearly shows which component of the corresponding second order susceptibility tensor is larger. For NC1′ we observe that $\left|\chi^{(2)}_{xxx}\right| > \left|\chi^{(2)}_{xyy}\right|$, while for NC2′ $\left|\chi^{(2)}_{xxx}\right| \approx \left|\chi^{(2)}_{xyy}\right|$. Although our model qualitatively describes angular properties of the observed results, one important piece is still missing, namely the experimental angular diagrams show clear asymmetry of the LA polarized SH signal (Figure 4).

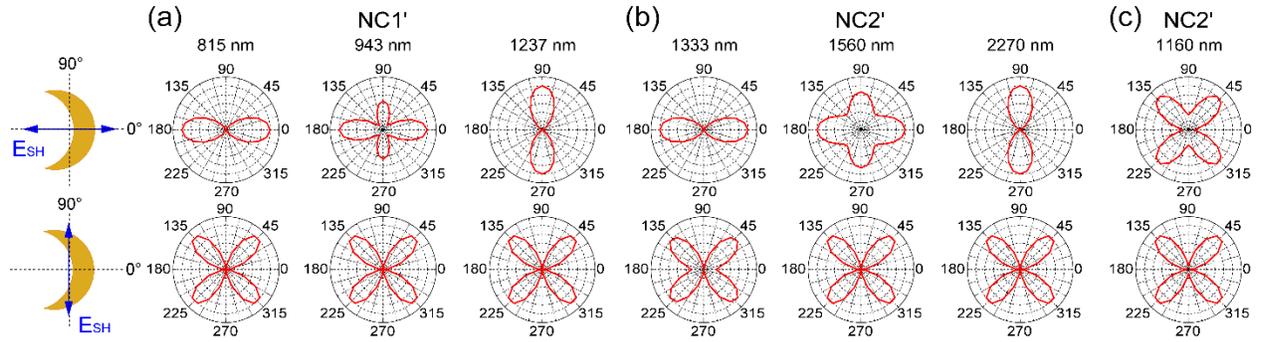

**Figure 6.** Incident polarization angle dependence of SH signals calculated for NC1′ (a) and NC2′ (b and c) at different pump wavelengths. The polarization angle is defined relative to the SA of the NC. The upper (lower) row shows SH signals polarized along the SA (LA) of the NC.

In using nanosphere template lithography to fabricate an NC array, not every structure is *perfectly* symmetric. Distorted structure breaks the symmetry, which makes the incident polarization angle dependence of SH not theoretically reproduced by eqs 1a and 1b. Also, the fitted coefficients of eq 2 would have no clear physical meaning in that case because their relationships with structural distortion cannot be straightforwardly predicted. Our model calculation, however, allows us to further investigate how the SH response is sensitive to the symmetry of nanocrescents. We consider two possible scenarios that may alter the symmetric angular patterns seen in Figure 6 to become asymmetric. Schematic setups that we have investigated are shown in Figure 7a,b. First, we rotate the NC counterclockwise by a small angle $\theta$ and alter the surface of the particle by adding extra-pixels. The latter is done for two sides of the particle (separated by the dashed line shown in Figure 7a) separately to ensure that the mirror symmetry is no longer maintained. Extra-pixels added to the upper part of the crescent are shown in blue and the pixels added to the lower part are shown in red. We note that the pixelization shown is schematic with overblown pixels. One can add more pixels to the upper inner part of the particle compared to the upper outer part and vice versa (similarly for the lower parts). It is important to emphasize that the described extra-pixelization may alter the volume of the particle if the total number of pixels is large, which in turn affects plasmonic properties. On the other hand, we must ensure that numerical convergence is achieved, i.e., when adding a small number of pixels, one should not observe any changes in the SHG signal. We thus need to find a delicate balance, in which, on one hand, the volume of the particle does not significantly depart from the volume that the symmetric particle has and on the other hand the surface is altered just enough to observe breaking of the mirror symmetry. After extensive numerical tests we found that the angular properties of SHG noticeably change for the total number of extra-pixels exceeding 1000 for the spatial resolution of 1.5 nm. Another way of breaking the mirror symmetry of the crescent is to superimpose an ellipse on a circle (rather than a circle on a circle) as depicted in Figure. 7b. In this case when the ellipse is shifted the resulting crescent no longer maintains mirror symmetry. When the ellipse is moved down the crescent becomes thicker at the top and thinner below the dashed line.

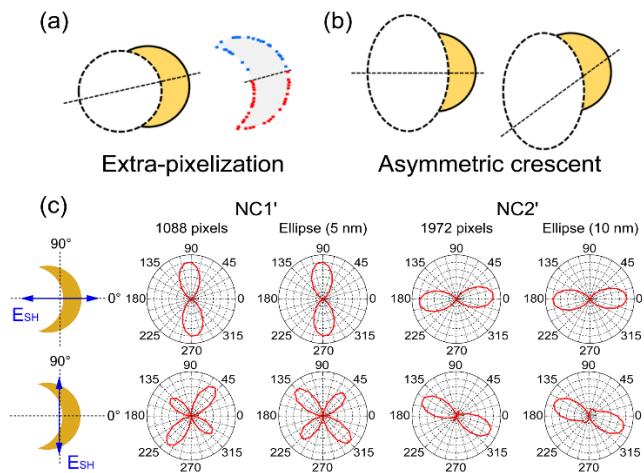

**Figure 7.** Broken symmetry of NC and its effects on the incident polarization angle dependence of calculated second harmonic intensity. Schematic diagrams in (a) and (b) show two ways the mirror symmetry of a NC can be broken. In extra-pixelization (a), the crescent is first rotated counterclockwise and then extra-pixels are added to the surface. Here blue and red color corresponds to pixels added to the upper and lower parts of the particle in non-symmetric fashion (each pixel does not have its mirror image on the other side of the dashed line). In asymmetrization (b), an ellipse is superimposed on a circle. When the ellipse is spatially shifted the resulting crescent does not have a mirror symmetry. Part (c) shows incident polarization angle dependence of SH signals calculated for NC1′ (first two columns) and NC2′ (last two columns) when extra-pixels are added to the surface and when ellipse-on-circle is used to generate the shape of the crescent. The upper (lower) row shows SH signal polarized along the SA (LA) of the NC. The rotation of NC1′ with extra-pixelization is by 3.4° and for NC2′ is 6.4°.

The results of both approaches that break mirror symmetry are summarized in Figure 7c. First and foremost, the obtained angular diagrams qualitatively show similar patterns to those seen experimentally with highly non-symmetric LA polarized SH signals. The total number of extra pixels added to NC1′ and NC2′ in Figure 7c is 1088 and 1972, correspondingly. There are two important observations: (1) the SA polarized signal rotates exactly by $\theta$ although it maintains symmetric two-lobe patterns; (2) the LA polarized signal is significantly altered by the symmetry breaking. Moreover, the observed asymmetry in the LA polarized SH signals for NC1′ (for which we examine the LA plasmonic mode contribution to SHG) shows another fascinating phenomenon, namely two smaller lobes (out of total four lobes) become shorter when the upper part of the non-symmetric crescent has less extra-pixels compared to the lower part. Furthermore, we found that extra-pixelization that alters the surface of the NC making it non-mirror symmetric can also explain which of four lobes in the LA SH become shorter/longer. For the calculations shown in Figure 7c, NC1′ has total 1088 extra pixels with more pixels placed at the outer surface (731) compared to the inner surface (357), while the number of extra pixels on the upper part of the particle (above dashed line in Figure 7a) is 527 with 561 pixels placed on the lower surface. However, when we break the symmetry using the ellipse-on-circle procedure, the four-lobe pattern exhibits completely opposite behavior. This is obviously because the resulting crescent has a thicker upper part as we discussed above. Even more dramatic changes occur for the NC2′ geometry, when we consider how the symmetry breaking influences the angular properties of SHG while pumping the SA plasmonic mode. The four-lobe pattern exhibits remarkable transformations with two lobes being nearly completely gone.

**E. Wavelength dependence of SH response from NC arrays.** Figure 8a shows the incident wavelength dependence of SH intensities (left axis) measured for NC1 (top, black circles) and NC2 (bottom, black squares). These experiments employed the same polarization configuration of the incident near-IR and detected SH light as shown in Figure S2a: excitation along the LA and detection along the SA for NC1, and excitation and detection along the SA for NC2. For comparison, the linear absorbance spectra collected with the light polarized along the LA (red) and SA (blue) are plotted (right axis) as well. Clearly, the wavelength dependence of the SH light intensity does not follow the spectral shapes of the dipole LSPRs of Au NC. For NC1, the SH intensity exhibits a dip at around 1275 nm, and the maximum response occurs at ~1375 nm. For NC2, the SH intensity monotonically increases with decreasing the incident wavelength with a small bump at 1425 nm.

To elucidate the effects of higher-order plasmon resonances on the SH enhancement, it is useful to compare the excitation wavelength dependence of SH with plasmon band shapes at half of the incident wavelength. Figure 8b shows the SH intensities together with the absorbance spectra in the wavelength range from 400 to 1000 nm. For NC1, the dependence is quite different from the absorbance spectrum observed with the light polarized along SA (blue curve) in this wavelength range, but it shows a similar trace to the "merged" shape of the LA dipole band (red curve in Figure 8a) and the dipole and multipole band (blue curve in Figure 8b). Although the SH trace also shows a similar shape to the LA quadrupole and multipole bands (red curve in Figure 8b), we do not expect these LSPR bands to contribute to SH enhancement because they are not covered by the excitation wavelength range. For NC2, the trend of SH response is similar to the right-hand-side of the multipole bands in the range from 600 to 700 nm.

The observed wavelength dependence of the SH intensity (Figure 8) implies that the SH response of Au NC array is not simply determined by the dipole LSPR bands. Intriguingly, the experimental data for NC1 resembles that observed by Niesler et al. for a split-ring-resonator array with a similar polarization dependent LSPR band profile.[14] Although they concluded that the role of LSPR at the SH wavelength was merely to reabsorb the induced SH signal, this is not the case for the Au NC array studied here. For example, NC2 has the largest SH intensity at 1200 nm, while the multipole band also shows maximum absorbance at half of that wavelength. If the SH response is determined solely by $L(\omega)$, then the peak is expected to be located at 1277 nm. It should be mentioned that we did not apply the anharmonic oscillator model to calculate the SH wavelength dependence based on the linear susceptibilities. This model can be applied to nanostructures with the LSPR located only at the SH wavelength to reproduce measured SH spectra.[17,24] In these systems, the resonance does indeed play an important role to govern the SH response. However, it has been argued that the model predicts incorrect behavior of induced SH for more complicated structures,[16] and it is not a suitable model when higher-order modes are involved in the SH generation process.[33]

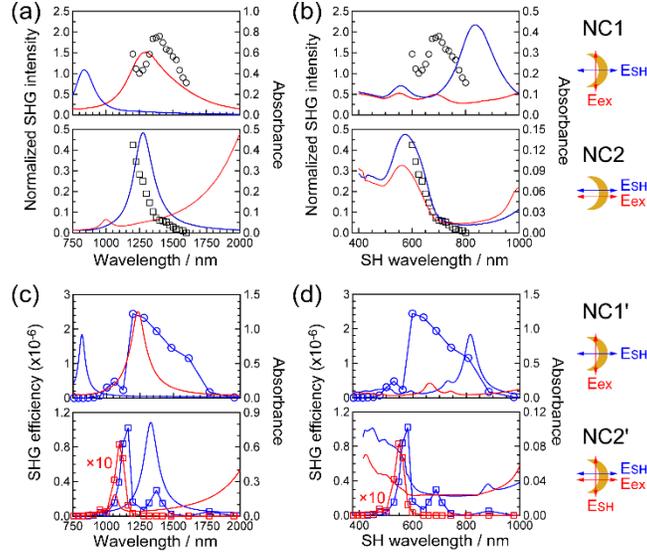

**Figure 8.** Measured and calculated excitation wavelength dependence of SH signals of Au NC arrays. Open circles and squares in panel (a) show integrated spectral intensity of NC1 and NC2, respectively, normalized with that of z-cut quartz. The polarization directions of the excitation pulse and the SH signal are indicated by a red and a blue arrow, respectively. Red and blue lines represent the absorbance spectra measured with the incident light polarized along the LA and SA of the NC, respectively. In panel (b), the same SH data set is plotted against one half of the incident wavelength together with the absorbance spectra. The top panel of (c) shows SHG efficiency of NC1′ calculated with the pump polarized along the LA and SH signal polarized along the SA (blue circles). In the same panel calculated linear absorbance is plotted, where the red line indicates absorbance obtained when the excitation is polarized along the LA, and the blue line with the incident field polarized along the SA. The bottom panel of (c) shows the results for NC2′ obtained with the pump polarized along the SA. Both polarizations of SHG efficiency are plotted (SA is shown as blue squares, and LA is shown as red squares after multiplying by 10). In panel (d), the same data set is plotted against one half of the incident wavelength together with the calculated absorbance spectra.

**F. Calculated wavelength dependence of SH signal.** Figure 8c exhibits the calculated SHG efficiency of NC1′ and NC2′ as a function of the pump wavelength. Blue circles (top) show the result for the SA polarized SH signal when NC1′ is pumped along its LA. Here we plot SHG efficiency and linear absorbance (blue and red lines) as functions of the incident wavelength. The SH response peaks at the LA plasmon mode at 1237 nm (Figure 8c) indicating that the dipole plasmon is a major reason for enhancing SH efficiency. At the same time, the SH response is broad and intense in the wavelength range > 1237 nm. It appears that both the SA dipole band (blue in Figure 8d) and the LA quadrupole/multipole bands (red in Figure 8d) can contribute to enhance SH response in this range. More interesting results are obtained for NC2′ (bottom).

The system is excited along the SA direction and the SHG efficiency along the same axis is plotted (blue squares). Additionally, the linear absorbance calculated for the SA and the LA polarized incident pump is overlapped in the same panel. Firstly, we observe that the SA polarized SH has one extremum near the peak wavelength of the corresponding dipole plasmon resonance. Secondly, we see that the efficiency curve has a remarkably strong extremum at ~1160 nm. Although it is near the peak of LA quadrupole resonance (1065 nm), we do not expect the LA quadrupole band to contribute to the enhancement because the incident light is polarized along the SA. Further observation shows that this SH peak is at the edge of a broad multipole resonance as shown in Figure 8d. The shape of the SH trace is similar to the merged shape of the SA dipole band (blue curve in Figure 8c) and the broad multipole band (blue curve in Figure 8d).

The results of our numerical calculation indicate that the SHG efficiency is affected not only by the field enhancement at the fundamental wavelength, but also at the SH wavelength. Depending on the size of the NC and the excitation wavelength, different LSPRs will couple with the induced SH field and render additional enhancement of signal intensity. For example, we found for NC2′ that the highest SHG efficiency is obtained at ~1160 nm, which is about 170 nm off from the peak of the SA dipole resonance. To further demonstrate that the multipole plasmon enhances the SHG at 1160 nm, we performed simulations for NC2′ by pumping the NC at 1160 nm and varying the in-plane incident field polarization. The results are shown in Figure 6c where a four-lobe pattern is observed for SA polarized SHG. This pattern indicates that $P_x^{(2)}(\omega) \sim \chi_{xxy}^{(2)} \sin\theta \cos\theta$ and $\chi_{xxy}^{(2)}$ is non-zero at this wavelength. This tensor element is dipole forbidden and hence the multipole LSPR at the SH wavelength has enhanced the induced field. The calculated multipole LSPR for NC2′ has a very broad line shape below 600 nm. However, the field enhancement by the SA dipole resonance is much weaker for the fundamental light below 1000 nm, and hence the SHG efficiency would not monotonically increase at shorter wavelengths. To further illustrate the effects of higher order LSPR, we calculated the SH efficiency of NC2′ along the LA (red squares in Figure 8c,d). The efficiency of the LA polarized SH is about 10 times lower than that of the SA polarized SH because the LA multipole is much weaker than the SA multipole resonance. The peak of the LA polarized SH is red shifted from that of SA polarized SH toward the LA quadrupole band at 1065 nm, suggesting the contribution of the LA quadrupole band to SHG.

## V. Conclusions

We have experimentally and theoretically investigated the wavelength and polarization dependent SH responses of two different Au NC arrays. Our data showed that the SH field induced in these nanoantennas cannot be described by the conventional formalism of second-order nonlinear polarization based on structural symmetry. The SH responses of NCs exhibit strong wavelength dependence over the scanned range from 1200 to 1600 nm, which covers the major part of the LA/SA dipole resonance bands. Our results suggest that LSPR plasmons at the SH wavelength also play significant roles, such as enhancement of the emitting SH field, and they have to be properly modeled for quantitative comparisons between the experiment and calculated results. The incident polarization angle dependence of LA polarized SH signals exhibits asymmetry that is extremely sensitive to nanoscale variations in the NC's shape. The polarization dependence of SH intensity for NC2 significantly depends on the wavelength of light used for measurement, much more so than NC1. It may be due to the fact that the SA dipole band is overlapping with the wing of the LA dipole band and the presence of multipole bands in the SH wavelength range. Further improvement in modeling is needed to quantitatively include the relative contributions from all these bands and their interplay.

**Supporting Information Available:** Schematic diagram of experimental setup of SH spectroscopy measurements, SH spectra, incident power and polarization dependence of SH signal, and calculated electric field and intensity distribution maps of LSPR and SH of Au NC arrays. This material is available free of charge via the Internet at http://pubs.acs.org

**Acknowledgements**

This research work was supported by grants from the US National Science Foundation (CHE-1414466 and CHE-0802913) to N.-H. G., J. S.-P., and G. C. S., and the Air Force Office of Scientific Research (FA9550-19-1-0009) to M. S. Secondary instrumentation and personnel supports were, respectively, from US NSF CHE-1310693 and CHE-1905395 to N.-H. G. C. A. L. also acknowledges support from an NSF Integrative Graduate Education and Research Traineeship (DGE-0903715). The authors also acknowledge computational support through Department of Defense High Performance Computing Modernization Program, and the use of UC Irvine Laser Spectroscopy Facility. The research made use of University of Utah shared facilities of Micron Technology Foundation, Inc. Microscopy Suite sponsored by the College of Engineering, Health Sciences Center, Office of the Vice President for Research, and the Utah Science Technology and Research (USTAR) initiative of the State of Utah and the University of Utah USTAR shared facilities supported, in part, by the NSF MRSEC Program (DMR-1121252).

**TOC Graphic**

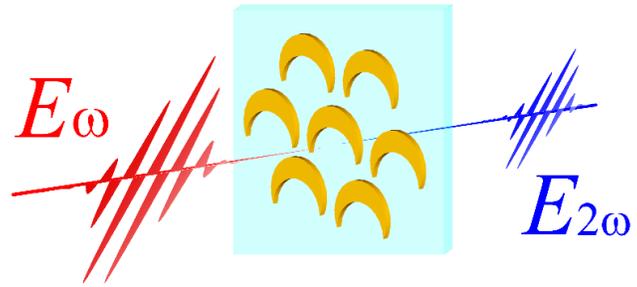

# Wavelength and Polarization Dependence of Second Harmonic Responses from Gold Nanocrescent Arrays

## Supporting Information

Hiroaki Maekawa,[1] Elena Drobnyh,[2] Cady A. Lancaster,[3] Nicolas Large,[4] George C. Schatz,[5] Jennifer S. Shumaker-Parry,[3] Maxim Sukharev,[2,6,*] and Nien-Hui Ge[1,*]

[1] Department of Chemistry, University of California at Irvine, Irvine, California 92697-2015, USA

[2] Department of Physics, Arizona State University, Tempe, Arizona 85287, USA

[3] Department of Chemistry, University of Utah, Salt Lake City, Utah 84112, USA

[4] Department of Physics and Astronomy, University of Texas at San Antonio, San Antonio, Texas 78249, USA

[5] Department of Chemistry, Northwestern University, Evanston, Illinois 60208-3113, USA

[6] College of Integrative Sciences and Arts, Arizona State University, Mesa, Arizona 85212, USA

* To whom correspondence should be addressed.

Maxim Sukharev: Maxim.Sukharev@asu.edu

Nien-Hui Ge: nhge@uci.edu

**Experimental setup of second-harmonic (SH) spectroscopy measurement**

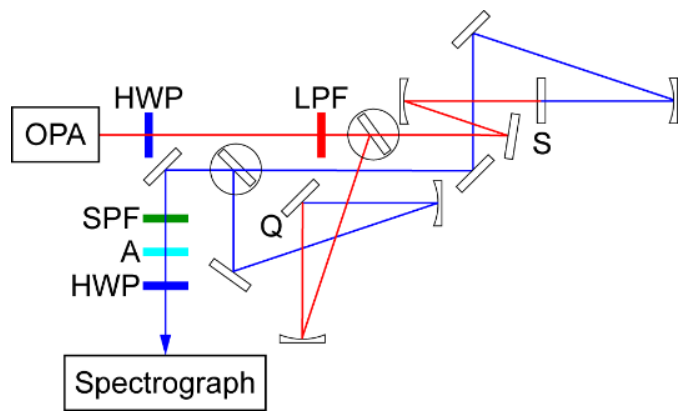

**Figure S1.** A schematic diagram of SH spectrometer. OPA: optical parametric amplifier; HWP: λ/2 wave plate; LPF: long-pass filter; S: sample of NC array; Q: z-cut quartz plate; SPF: short-pass filter; A: analyzer for SH signal. Two circles represent mirrors in flipper mounts to switch the optical paths between the sample and reference measurements.

**Experimental results of SH spectroscopy measurement for Au nanocrescent (NC) arrays**

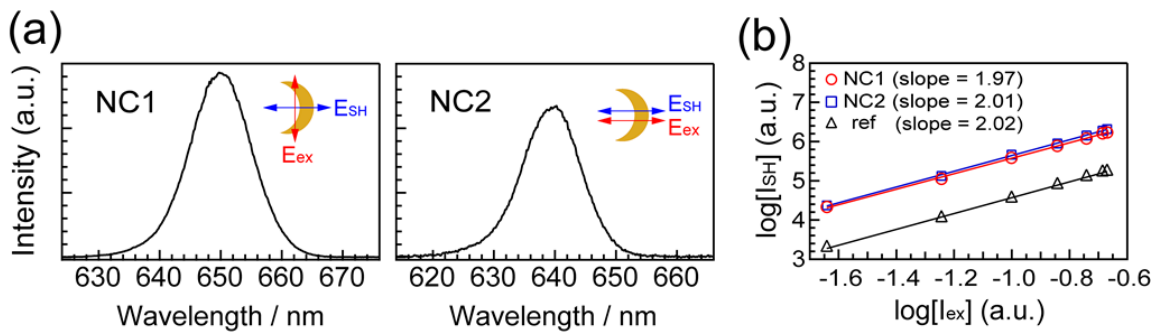

**Figure S2.** (a) Spectra of SH signal measured for NC1 with the excitation pulse centered at 1300 nm (left) and NC2 centered at 1280 nm (right). The polarization direction of the excitation pulse and the SH signal is shown by a red and a blue arrow in each panel, respectively. (b) Double logarithmic plot of incident power dependence of SH signal: (red circle) NC1; (blue square) NC2; and (black triangle) z-cut quartz used as a reference. Solid lines represent the results of linear least squares fitting to the experimental data.

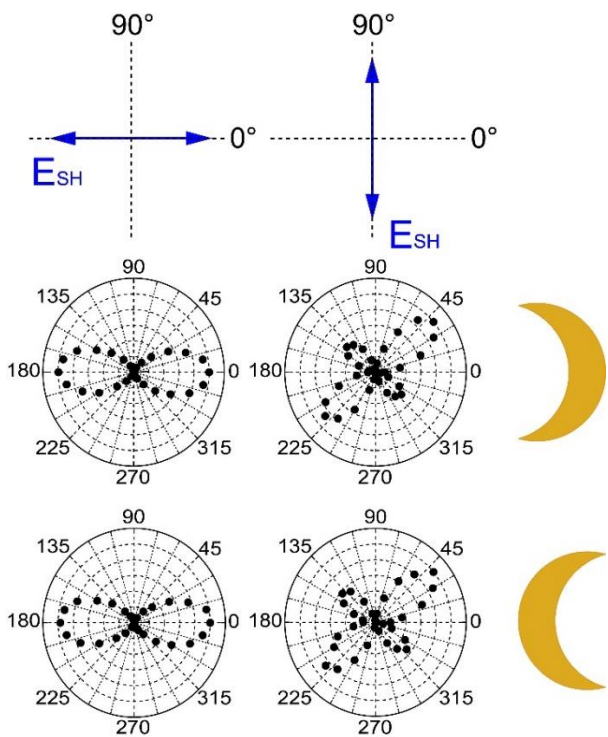

**Figure S3.** Polarization dependence of SH signals measured for NC2 at the excitation wavelength of 1280 nm. Rotating the samples by 180° on the sample plane does not affect the SH response.

All field components and intensity distributions are calculated on the input side 15 nm above the surface of NC. Distributions are normalized with respect to the pump amplitude.

**Electric field and intensity distributions of dipole LSPRs calculated for Au NC arrays NC1′ and NC2′**

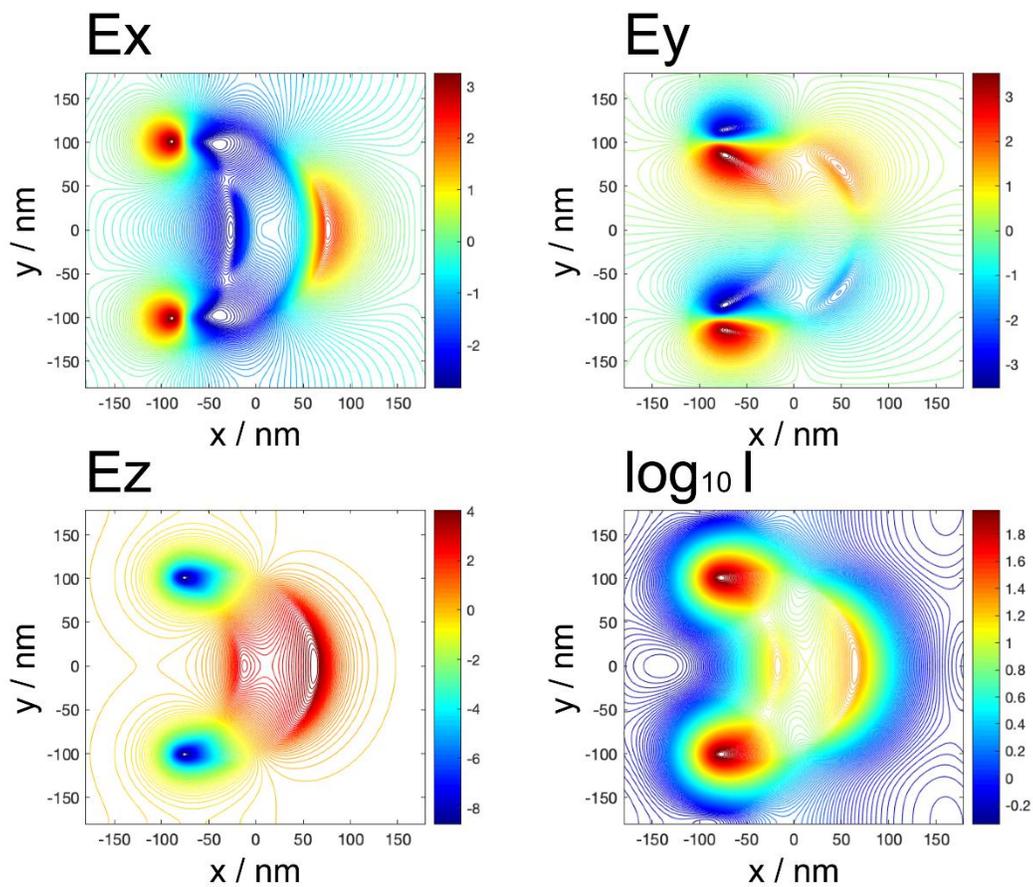

**Figure S4.** The amplitude of electric field ($E_i$, $i$ = x, y, z) and intensity (I) distributions calculated at the peak wavelength of the short-axis dipole LSPR (815 nm) for NC1′: (top left) $E_x$; (top right) $E_y$; (bottom left) $E_z$; (bottom right) $\log_{10}I$. Units of electric field correspond to the enhancement relative to the incident field.

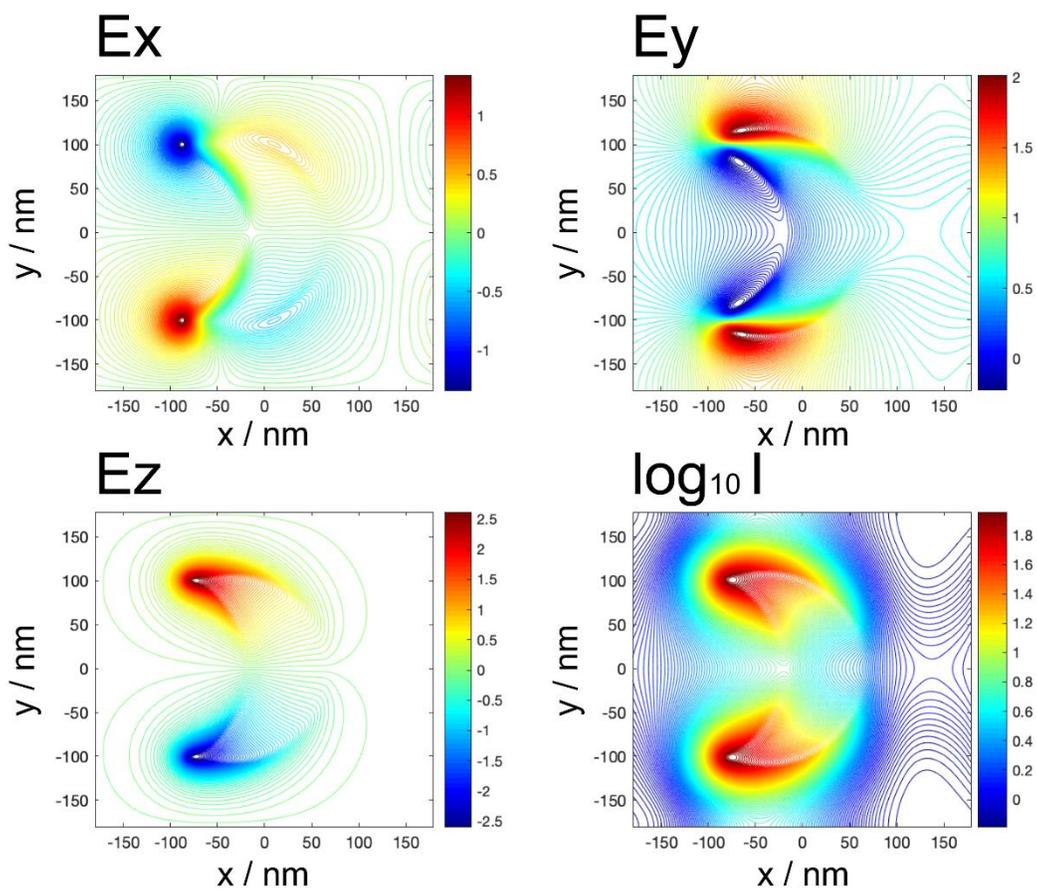

**Figure S5.** The amplitude of electric field ($E_i$, $i$ = x, y, z) and intensity (I) distributions calculated at the peak wavelength of the long-axis dipole LSPR (1237 nm) for NC1′: (top left) $E_x$; (top right) $E_y$; (bottom left) $E_z$; (bottom right) $\log_{10}I$. Units of electric field correspond to the enhancement relative to the incident field.

S5

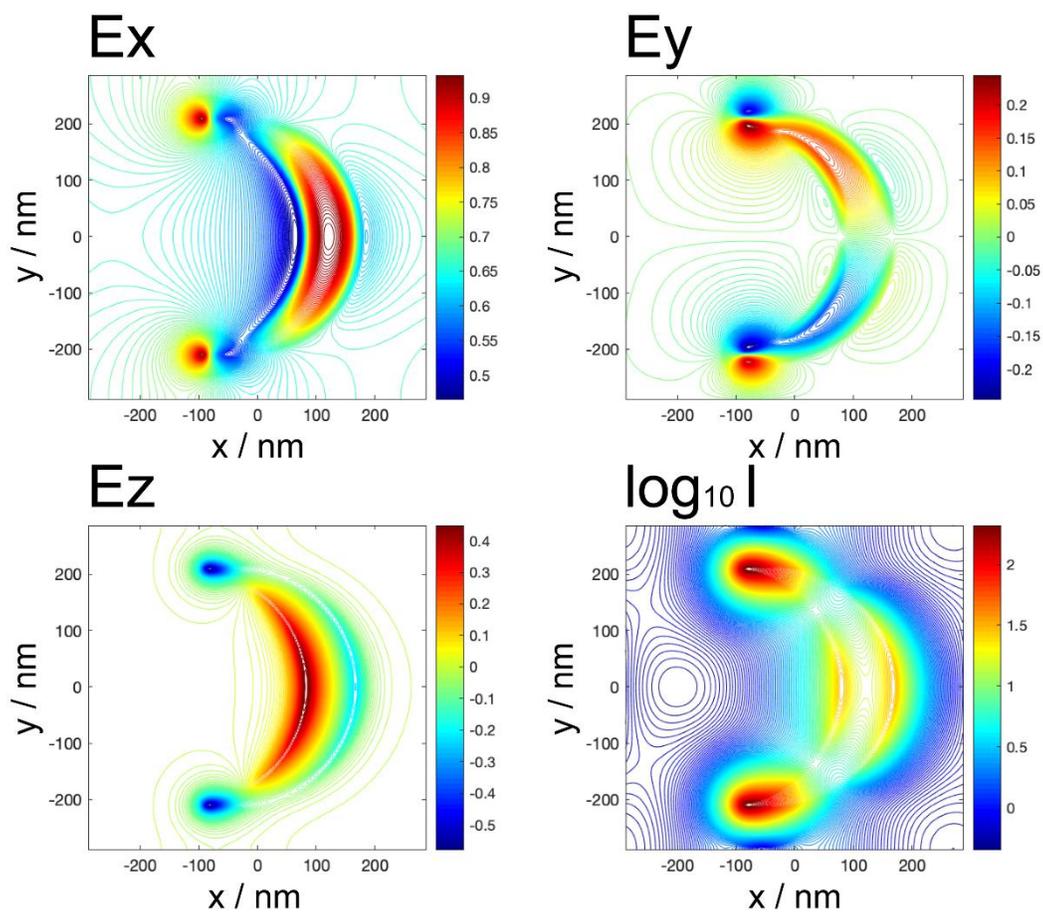

**Figure S6.** The amplitude of electric field (E$_i$, $i$ = x, y, z) and intensity (I) distributions calculated at the peak wavelength of the short-axis dipole LSPR (1333 nm) for NC2′: (top left) E$_x$; (top right) E$_y$; (bottom left) E$_z$; (bottom right) log$_{10}$I. Units of electric field correspond to the enhancement relative to the incident field.

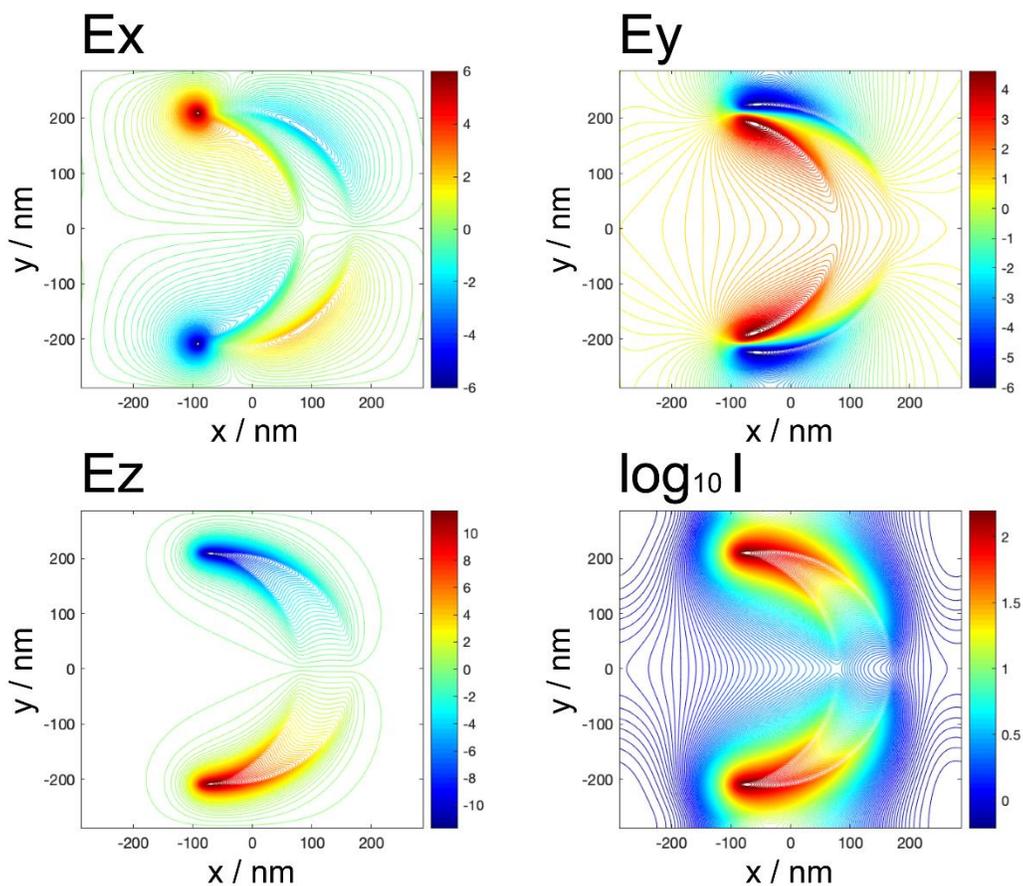

**Figure S7.** The amplitude of electric field ($E_i$, $i$ = x, y, z) and intensity (I) distributions calculated at the peak wavelength of the long-axis dipole LSPR (2270 nm) for NC2′: (top left) $E_x$; (top right) $E_y$; (bottom left) $E_z$; (bottom right) $\log_{10}I$. Units of electric field correspond to the enhancement relative to the incident field.

**Second-harmonic electric field and intensity distributions calculated for Au NC arrays NC1′ and NC2′**

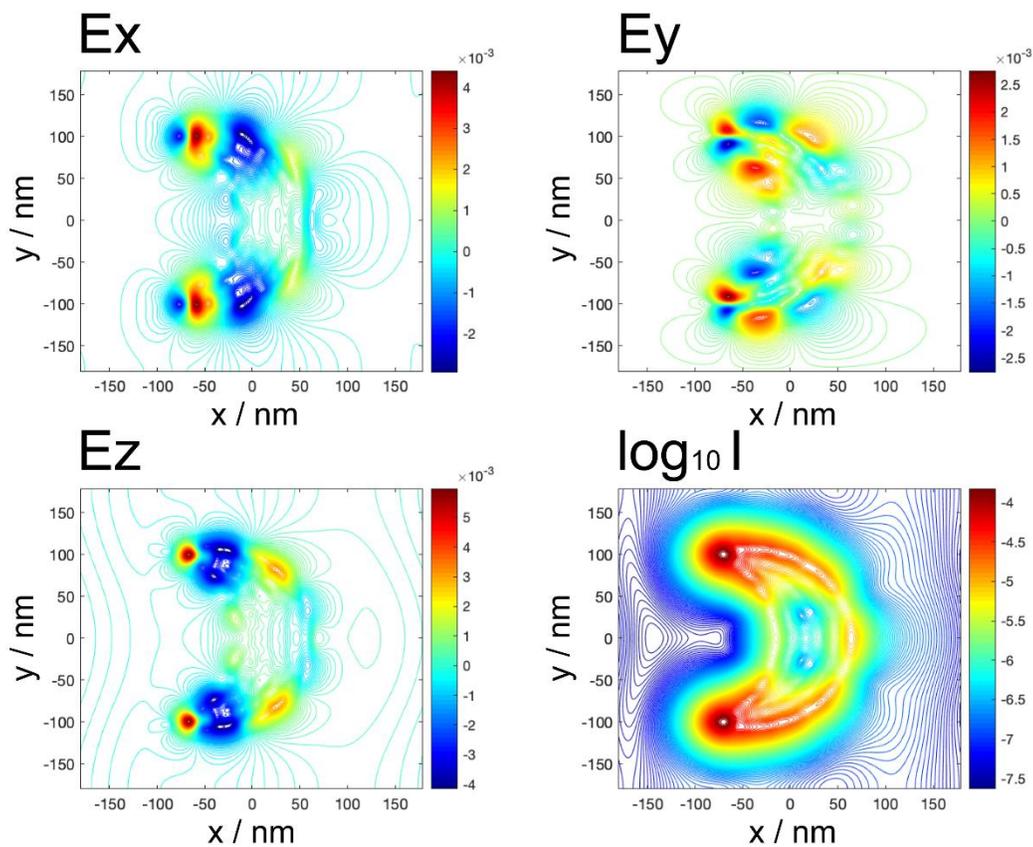

**Figure S8.** The amplitude of SH electric field ($E_i$, $i$ = x, y, z) and intensity (I) distributions calculated with the excitation at peak wavelength of the short-axis dipole LSPR (815 nm) for NC1′: (top left) $E_x$; (top right) $E_y$; (bottom left) $E_z$; (bottom right) $\log_{10}I$. Units of electric field correspond to the enhancement relative to the pump.

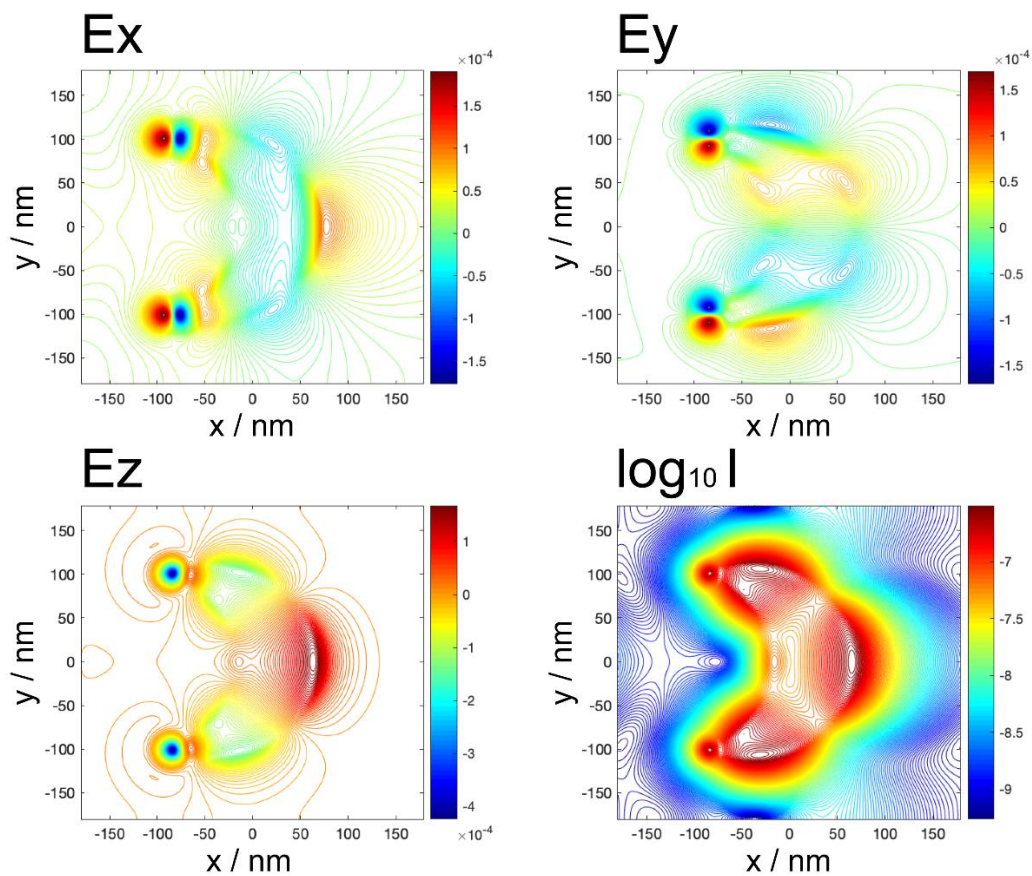

**Figure S9.** The amplitude of SH electric field ($E_i$, $i$ = x, y, z) and intensity (I) distributions calculated with the excitation at peak wavelength of the long-axis dipole LSPR (1237 nm) for NC1′: (top left) $E_x$; (top right) $E_y$; (bottom left) $E_z$; (bottom right) $\log_{10}I$. Units of electric field correspond to the enhancement relative to the pump.

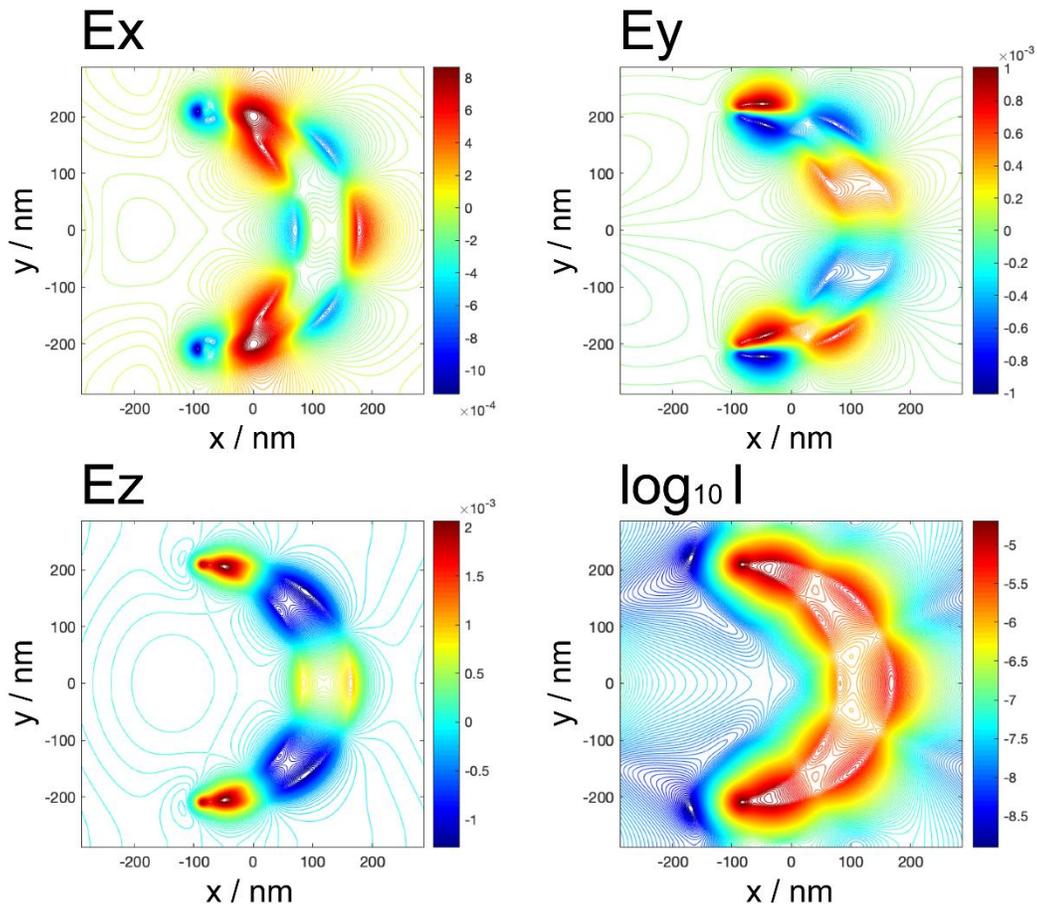

**Figure S10.** The amplitude of SH electric field ($E_i$, $i$ = x, y, z) and intensity (I) distributions calculated with the excitation at peak wavelength of the short-axis dipole LSPR (1333 nm) for NC2′: (top left) $E_x$; (top right) $E_y$; (bottom left) $E_z$; (bottom right) $\log_{10}I$. Units of electric field correspond to the enhancement relative to the pump.

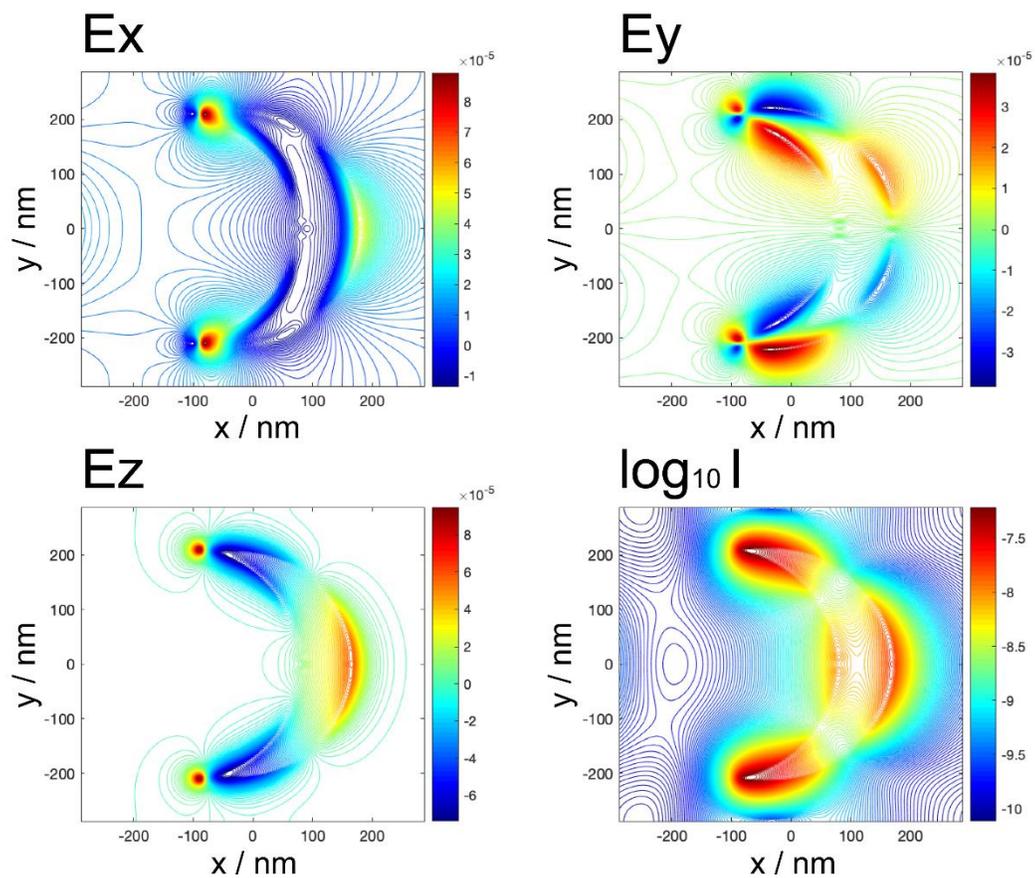

**Figure S11.** The amplitude of SH electric field ($E_i$, $i$ = x, y, z) and intensity (I) distributions calculated with the excitation at peak wavelength of the long-axis dipole LSPR (2270 nm) for NC2′: (top left) $E_x$; (top right) $E_y$; (bottom left) $E_z$; (bottom right) $\log_{10}I$. Units of electric field correspond to the enhancement relative to the pump.

**Electric field and intensity distributions of quadrupole LSPRs calculated for Au NC array NC2′**

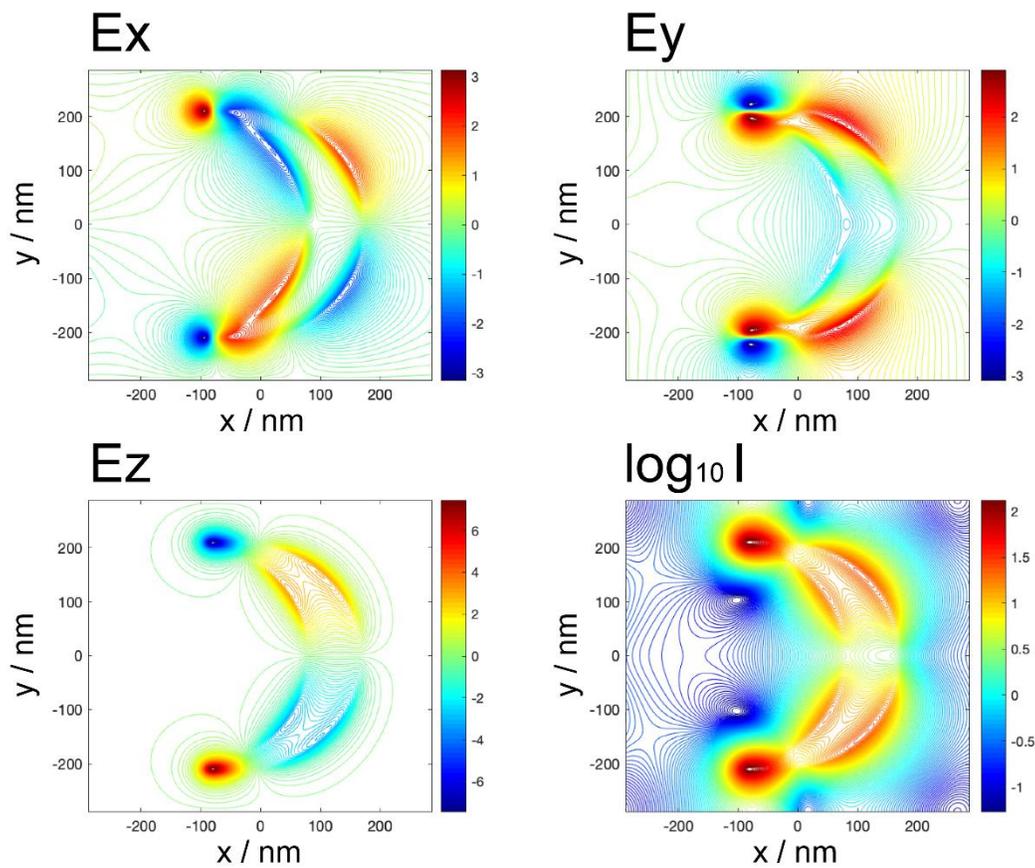

**Figure S12.** The amplitude of electric field ($E_i$, $i$ = x, y, z) and intensity (I) distributions calculated at the peak wavelength of the quadrupole LSPR (1064 nm) for NC2′: (top left) $E_x$; (top right) $E_y$; (bottom left) $E_z$; (bottom right) $\log_{10}I$. The incident field is polarized along the long axis of NC. Units of electric field correspond to the enhancement relative to the incident field.

**Second-harmonic electric field and intensity distributions calculated for Au NC arrays NC2′**

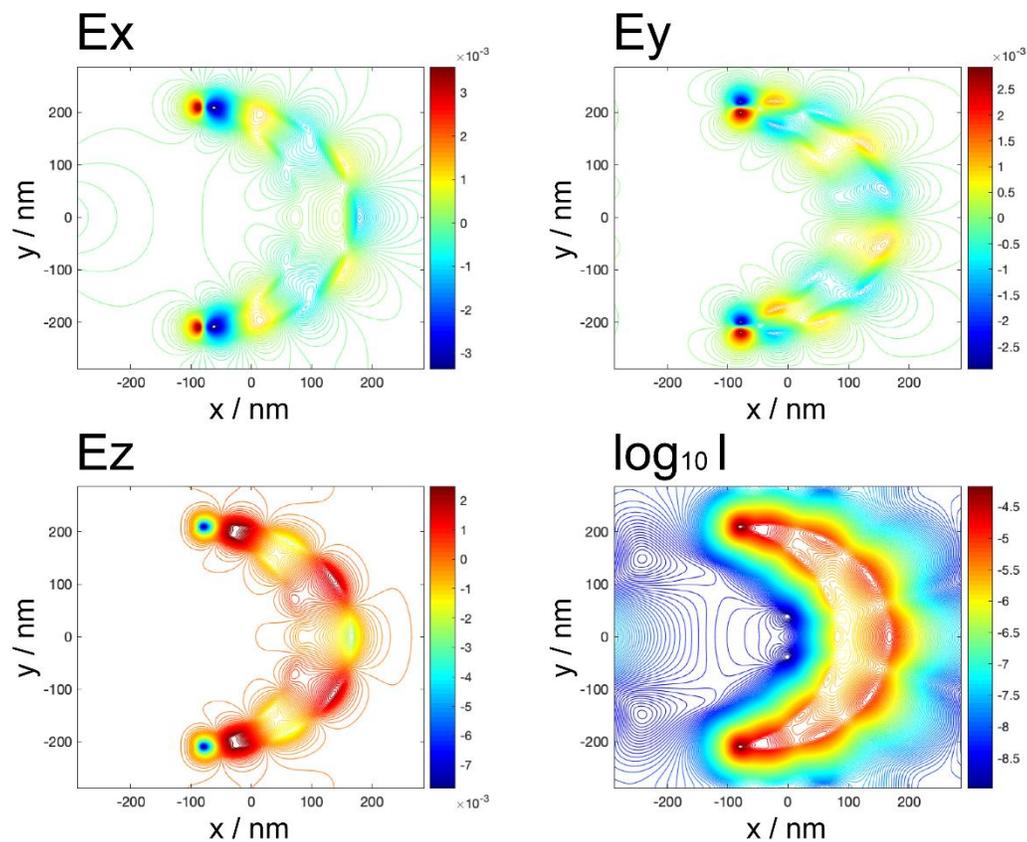

**Figure S13.** The amplitude of SH electric field ($E_i$, $i$ = x, y, z) and intensity (I) distributions calculated with excitation at the peak wavelength of the quadrupole LSPR (1064 nm) for NC2′: (top left) $E_x$; (top right) $E_y$; (bottom left) $E_z$; (bottom right) $\log_{10}I$. The incident field is polarized along the short axis of NC. Units of electric field correspond to the enhancement relative to the pump.

**Electric field and intensity distributions of multipole LSPRs calculated for Au NC arrays NC1′ and NC2′**

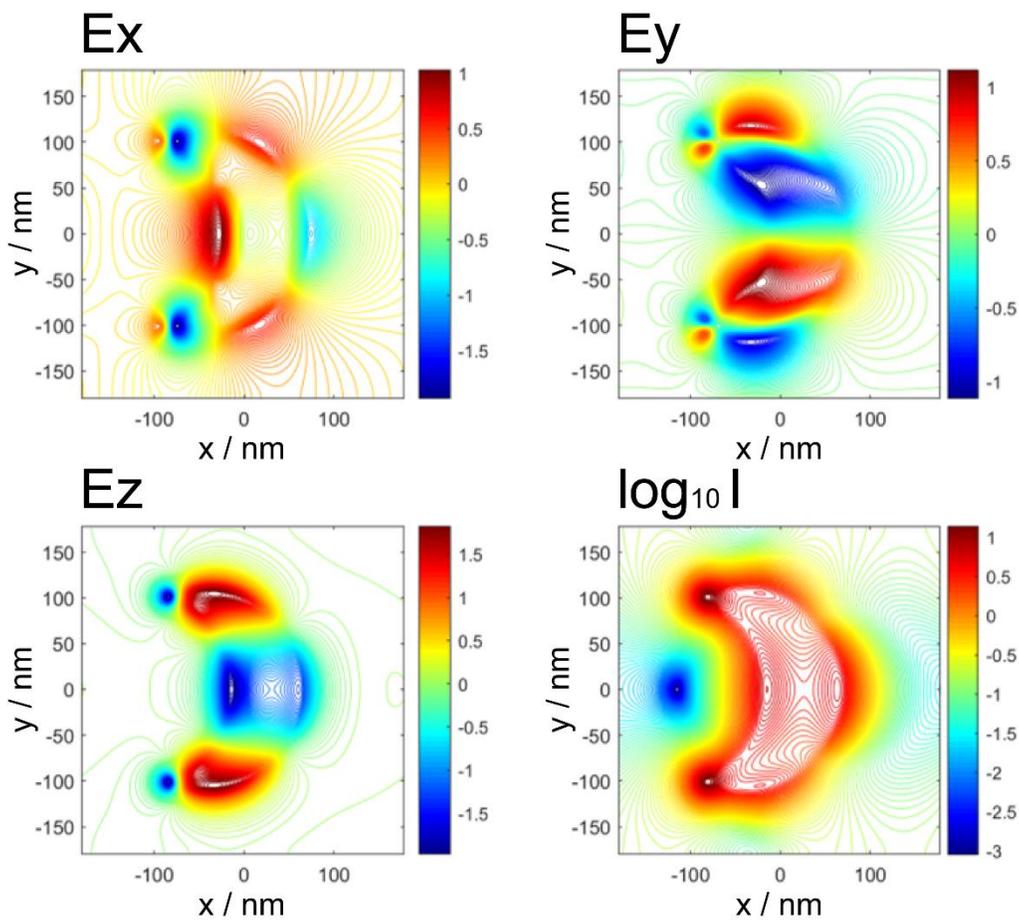

**Figure S14.** The amplitude of electric field ($E_i$, $i$ = x, y, z) and intensity (I) distributions calculated at the peak wavelength of the short-axis multipole LSPR (592 nm) for NC1′: (top left) $E_x$; (top right) $E_y$; (bottom left) $E_z$; (bottom right) $\log_{10}I$. Units of electric field correspond to the enhancement relative to the incident field.

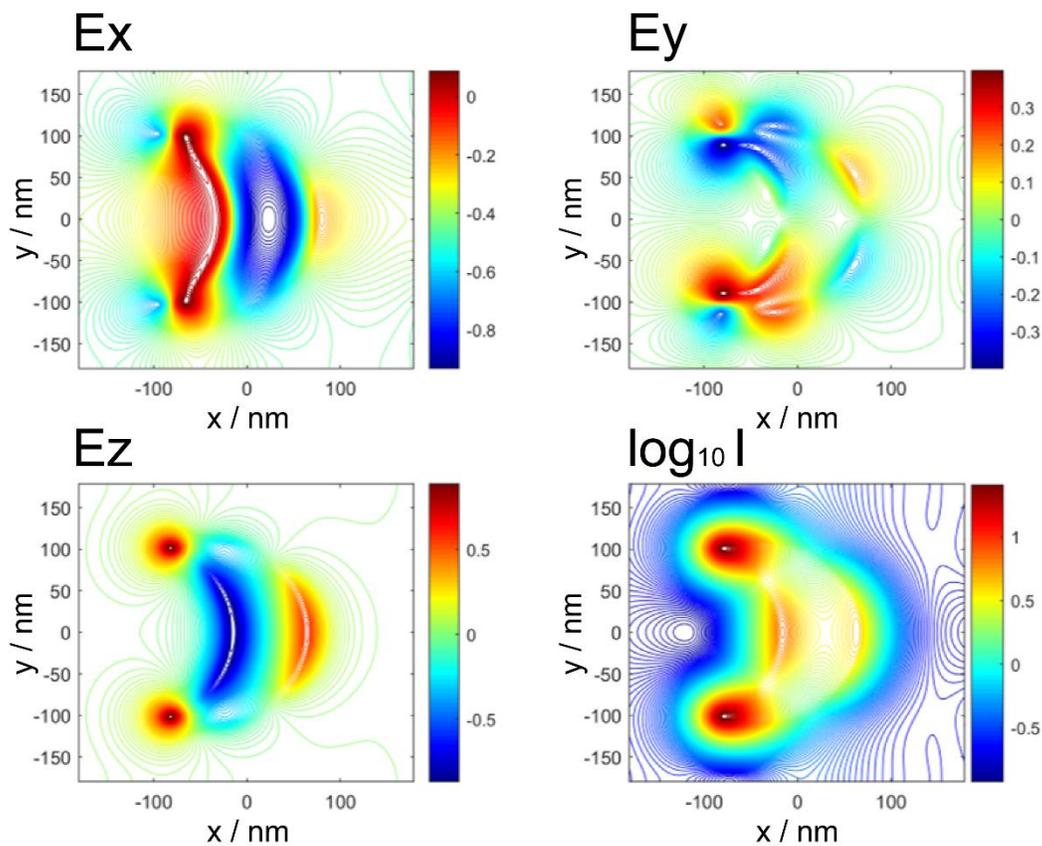

**Figure S15.** The amplitude of electric field ($E_i$, $i$ = x, y, z) and intensity (I) distributions calculated at the peak wavelength of the short-axis multipole LSPR (732 nm) for NC1′: (top left) $E_x$; (top right) $E_y$; (bottom left) $E_z$; (bottom right) $\log_{10}I$. Units of electric field correspond to the enhancement relative to the pump.

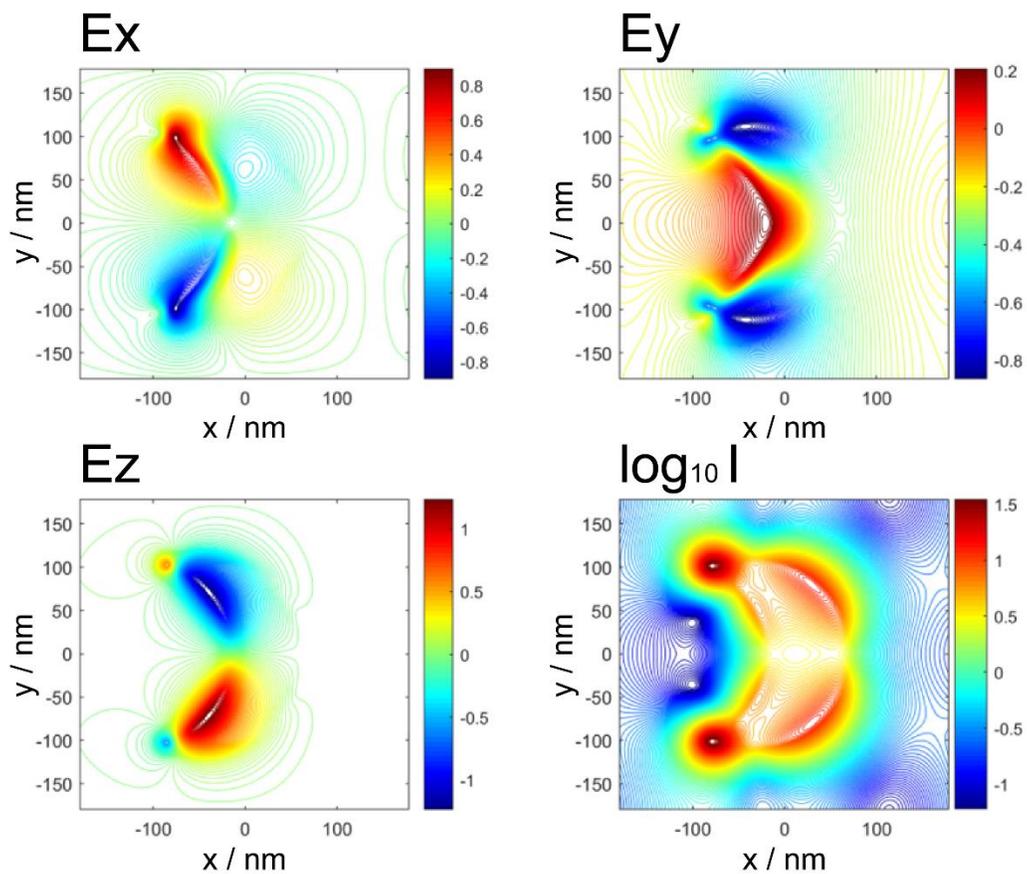

**Figure S16.** The amplitude of electric field ($E_i$, $i$ = x, y, z) and intensity (I) distributions calculated at the peak wavelength of the long-axis multipole LSPR (661 nm) for NC1′: (top left) $E_x$; (top right) $E_y$; (bottom left) $E_z$; (bottom right) $\log_{10}I$. Units of electric field correspond to the enhancement relative to the pump.

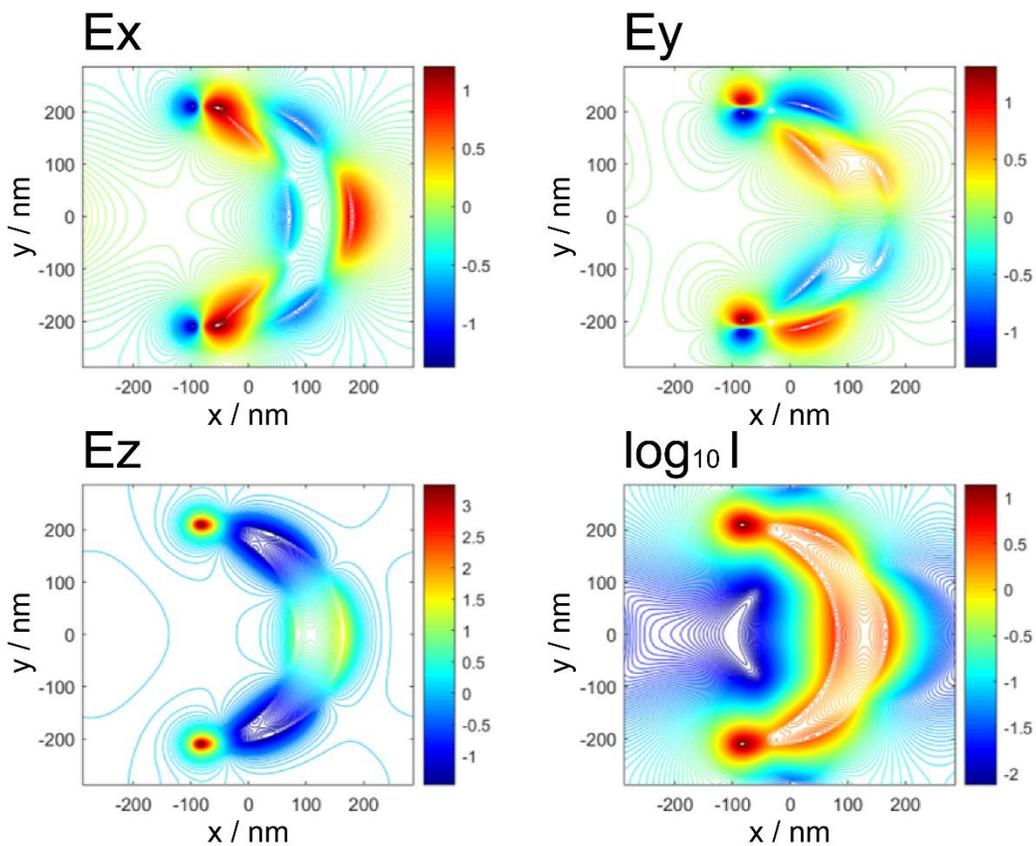

**Figure S18.** The amplitude of electric field ($E_i$, $i$ = x, y, z) and intensity (I) distributions calculated at the peak wavelength of the short-axis multipole LSPR (882 nm) for NC2′: (top left) $E_x$; (top right) $E_y$; (bottom left) $E_z$; (bottom right) $\log_{10}I$. Units of electric field correspond to the enhancement relative to the pump.

S17